\documentclass[pre,showpacs,amsmath,amssymb]{revtex4}
\usepackage{graphicx}
\usepackage{bm}
\begin{document}

\title{Spatial survival probability for one-dimensional fluctuating
interfaces in the steady state}

\author{ Satya N. Majumdar}
\affiliation{Laboratoire de Physique Theorique et Modeles Statistiques,
Universite Paris-Sud,
Bat. 100,
91405 ORSAY cedex, France}
\author{ Chandan Dasgupta}
\affiliation{Centre for Condensed Matter Theory, Department of Physics, Indian
Institute of Science, Bangalore 560012, India
}

\begin{abstract}
We report numerical and analytic results for the spatial survival probability
for fluctuating one-dimensional interfaces with Edwards-Wilkinson or
Kardar-Parisi-Zhang dynamics in the steady state. 
Our numerical results are
obtained from analysis of steady-state profiles generated by 
integrating a spatially discretized form of the Edwards-Wilkinson equation to 
long times. We show that the survival probability exhibits scaling behavior
in its dependence on the system size and the ``sampling interval'' used in the
measurement for both ``steady-state'' and
``finite'' initial conditions. 
Analytic results for the scaling functions are obtained from a path-integral 
treatment of a formulation of the problem in terms of one-dimensional Brownian
motion. A ``deterministic approximation'' is used to obtain closed-form
expressions for survival probabilities from the formally exact 
analytic treatment. The resulting approximate analytic results provide a
fairly good description of the numerical data. 
\end{abstract}

\pacs{68.35.Ja, 05.40.Jc, 05.20.-y}

\maketitle
\date{today}

\section{Introduction}
\label{intro}

Temporal first-passage properties, expressed in terms of persistence and
survival probabilities~\cite{sm1}, have recently found many 
applications~\cite{th1,th2,th3,th4} in the
study of the dynamics of fluctuating interfaces. Experimental realizations
of one-dimensional (1d) fluctuating interfaces are provided by monatomic steps
on vicinal surfaces. Recent experimental studies~\cite{exp1,exp2,exp3,exp4} 
have shown that temporal 
persistence and survival probabilities for fluctuating steps can be 
measured, and that these probabilities provide a convenient way of 
characterizing their dynamics. The temporal persistence probability is
defined in this context 
as the probability that the height at a particular point of the
interface does not return to its {\it initial value} over a certain period of
time. For many simple models of interface dynamics, this probability exhibits
a power-law decay in time~\cite{th1,th2,th3}. This power-law behavior has been
confirmed in experiments~\cite{exp1,exp2,exp3,exp4}.
In contrast, a closely related quantity, the
temporal survival probability that measures the probability that the height 
does not return to its {\it average value} over a certain period of time, is
found, both theoretically~\cite{th4} and experimentally~\cite{exp3,exp4}
to decay exponentially at long times.  

In studies of fluctuating interfaces,
it is natural to consider spatial analogs of these 
temporal first-passage quantities, namely the {\it spatial persistence
and survival probabilities}. These probabilities have been studied 
analytically~\cite{sp1} and numerically~\cite{sp2} for several models of
interfacial dynamics. 
For (1+1)-dimensional interfaces, the
stochastic variable of interest is the ``height'' $h(x,t)$ that represents 
the position of the interface at point $x$ and time $t$. In this paper,
we consider the
interface profile $h(x,t_0)$ where the time $t_0$ is in the long-time, 
steady-state
regime, and for notational convenience, suppress the time argument $t_0$ 
of $h$ from now on. To define spatial persistence probabilities, let
$p(x_{0},x_{0}+x)$ be the probability that
the height $h(x)$ {\it does not} return to its ``initial'' value $h(x_0)$ 
at the point $x_0$ within a distance $x$ 
measured from $x_{0}$ along the interface. The ``steady-state'' spatial
persistence probability $P(x)$ is then defined as the average of
$p(x_{0},x_{0}+x)$ over {\it all} possible choices of the initial point $x_0$
in a steady-state configuration of the interface. A second persistence
probability $P_{FIC}(x)$, the so-called spatial persistence probability 
for ``finite
initial conditions'' (FIC)~\cite{sp1}, is defined as the average of 
$p(x_{0},x_{0}+x)$ over initial points $x_0$ at which both the height $h(x_0)$
and its spatial derivative $h^\prime(x_0)$ are finite. It was shown in
Refs.~\cite{sp1,sp2} that for several models of fluctuating interfaces,
both $P(x)$ and $P_{FIC}(x)$ exhibit power-law decay for large $x$, but the
exponents that describe this power-law behavior may be different 
in the two cases. Experimental measurements of the 
steady-state spatial persistence probability for interfaces
(combustion fronts in paper) believed to be described by the 
Kardar-Parisi-Zhang (KPZ) equation~\cite{kpz} have been reported 
recently~\cite{exp_px}. The behavior of $P_{FIC}(x)$ was not
investigated in this work.

In analogy with the temporal case, the
{\it spatial survival probabilities} are defined in terms of the probability
$p^\prime(x_0,x_0+x)$ that the interface height between points $x_0$
and $x_0+x$ does not cross
its {\it average value} $\bar{h}$ (rather than the
initial value $h(x_0)$). The steady-state and FIC
spatial survival probabilities, $Q(x)$ and $Q_{FIC}(x)$, respectively, are 
then obtained by averaging $p^\prime(x_0,x_0+x)$ over $x_0$ in the 
two different ways mentioned above:
in the first case, the average is done over all points $x_0$, while
in the second case, the average is performed over
only the points at which the height and its spatial derivative are finite.
Numerical results for the $x$-dependence of these two spatial survival
probabilities for 1d interfaces with KPZ and 
Edwards-Wilkinson (EW)~\cite{ew} dynamics were reported
in Ref.\cite{sp2}. It was found there that the
spatial dependence of $Q(x)$ is neither power-law, nor
exponential, while $Q_{FIC}(x)$ exhibits a power-law decay similar to 
that of the FIC persistence probability $P_{FIC}(x)$. While the power-law
behavior of $Q_{FIC}(x)$ was expected~\cite{sp1}, the $x$-dependence of
$Q(x)$ found in the numerical work of Ref.\cite{sp2} was not understood
theoretically.

In this paper, we present the results of a detailed numerical and analytic
study of the spatial survival probabilities for 1d interfaces
with EW dynamics in the steady state. The primary motivation for this study
is to develop a theoretical understanding of the numerical results reported
in Ref.\cite{sp2}. On a more general level, studies of temporal and
spatial first-passage properties of fluctuating interfaces
are believed to be important in understanding the role of 
thermal fluctuations of the edges of components 
in the stability of nano-scale devices.
The 1d EW equation is 
believed~\cite{bart} to describe thermal fluctuations of steps 
on a vicinal surface under experimental conditions for which the dominant
source of fluctuations is the attachment/detachment of atoms to/from 
the steps.
Since the statistics of height fluctuations in the steady state of the 
1d KPZ equation is the same as that for the 1d EW equation, our
results also apply to experimental realizations of 1d KPZ interfaces. 

The 1d EW equation has the form\cite{ew}
\begin{equation}
\frac{\partial h(x,t)}{\partial t} = \Gamma \frac{\partial^2 h(x,t)}
{\partial x^2} + \eta(x,t), \label{eweq}
\end{equation}
where $\Gamma$ is a kinetic parameter and $\eta(x,t)$ is a Gaussian random
noise with $\langle \eta(x,t) \rangle = 0$, $\langle \eta(x,t) \eta(x^\prime,
t^\prime) \rangle = 2D^\prime \delta(x-x^\prime) \delta(t-t^\prime)$, 
$D^\prime$ being a
measure of the strength of the noise. The parameters $\Gamma$ and $D^\prime$ 
should satisfy the fluctuation-dissipation relation if this equation is
supposed to describe equilibrium fluctuations (e.g. in the case of steps on
a vicinal surface), but they are independent parameters in a general
non-equilibrium situation. We consider a finite system of size $L$, so that
$0 \leq x \leq L$, and use periodic boundary conditions. 
It is easy to see that the spatial average $\bar{h}(t)
\equiv 1/L \int_0^L h(x,t) dx$ executes a simple random walk in time. In 
our calculations, we subtract the spatial average from the variables 
$h(x,t)$, so that from now on, it is implied that $h(x,t)$ represents the
height at point $x$ measured from the instantaneous spatial average. Thus,
$\bar{h}(t)$ is equal to zero by definition. As we shall see later, this
condition plays an important role in the analytic calculation. 

The spatial survival probabilities studied here are defined as follows.
Let $Q(x,L|h_0)$ denote the
probability that the steady state spatial profile of the interface
starting at $h_0$
at $x=0$ (and ending at $h_0$ at $x=L$, i.e., with periodic boundary condition)
does not cross $0$ upto a distance $x$ where $0\le x\le L$. Then, the
steady-state spatial
survival probability is given by
\begin{equation}
Q(x,L)= \int_{-\infty}^{\infty} Q(x,L|h_0) P_{st}(h_0,L) dh_0 ,\label{sseq}
\end{equation}
where $P_{st}(h_0,L)$ is the steady-state height distribution. The FIC
survival probability is defined as
\begin{equation}
Q_{FIC}(x,L,w)= \frac{\int_{-w}^{w} Q(x,L|h_0) P_{st}(h_0,L) dh_0}
{\int_{-w}^{w} P_{st}(h_0,L) dh_0} ,\label{ficeq}
\end{equation}
where $w \ll W(L)$, the steady-state width of the interface. In our numerical
work, steady-state profiles for systems of different $L$ are generated by 
integrating a spatially discretized form of the EW equation to long times,
and these profiles are used to calculate the survival probabilities. The
spatial discretization implies that there is a finite {\it sampling
interval} $\delta x$ that represents the spacing between successive points
at which the height variable is sampled in the calculation of the survival
probabilities. Clearly, the value of $\delta x$ must be an 
integral multiple of the
spacing of the spatial grid at which the height variable is defined. This
sampling interval $\delta x$ is analogous to the 
``sampling time'' that represents the interval between
two successive measurements of a stochastic process
in studies of temporal persistence. The fact that a finite value
of the ``sampling time" may modify the persistence properties
of a stationary stochastic process was first pointed out in
Ref. \cite{sm2}. In the context of fluctuating interfaces,
it is known from numerical~\cite{th3,th4} and 
experimental~\cite{exp4} studies that the temporal survival probability 
in interfaces exhibits a non-trivial dependence on the sampling time. In our numerical
study, we find that the spatial survival probabilities also depend on the
value of the sampling interval $\delta x$. The dependence of the survival
probability $Q$ on $x$, $L$ and $\delta x$
is found to be described by a scaling function of
$x/L$ and $\delta x/L$: $Q(x,L,\delta x) = f_d(x/L, \delta x/L)$. This is
similar to the scaling behavior found in
Ref.\cite{sp2} for spatial persistence probabilities. As shown there,
the dependence of the FIC survival 
probability $Q_{FIC}$ on $x$, $L$, $\delta x$ and $w$ is also 
described by a scaling function of $x/L$, $\delta x/L$, and $w/L^\alpha$
where $\alpha = 0.5$ is the exponent for the dependence of the steady-state
width $W(L)$ on $L$ [$W(L) \propto L^\alpha$].

In our analytic study, we consider the spatial survival probabilities when the
height variable is sampled continuously, i.e. the limit $\delta x \to 0$, and
calculate the scaling function $f(x/L) \equiv f_d(x/L,0)$. This calculation is
based on a mapping of the spatial statistics of a steady-state EW interface to
the temporal statistics of 1d Brownian motion. The requirement that the average
height $\bar{h}$ of the interface must vanish translates in this mapping 
to the constraint 
that the total area under the curve that represents the Brownian process in the
distance -- time plane must be zero. This ``zero-area'' constraint plays a very
important role in the analytic calculation -- the form of the scaling function
$f(u)$ depends crucially on whether this constraint is imposed in the analytic
treatment. Without this constraint, we can determine 
$f(u)$ exactly, but the form of the scaling function obtained this way 
differs drastically from
the numerical result. In particular, the scaling function does not go to 
zero as $u\to 1$, whereas
the numerically obtained  scaling function decreases rather fast to $0$ as $u$ 
approaches $1$. When we take into 
account the zero-area global constraint, determining the
scaling function $f(u)$ analytically becomes much more nontrivial.
We are able to set up an exact path integral technique that allows us, 
in principle,
to compute this function exactly in terms of some complicated integrals.
However, we cannot get an exact closed form expression for $f(u)$ 
to compare with the simulation data. We then
make a simple ``deterministic'' approximation that allows us to obtain a 
a closed form expression for $f(u)$ which we then compare
with the numerically obtained scaling function. The agreement is fairly good, 
given the drastic nature of the deterministic approximation. Our 
approximate analytic result for the FIC survival probability also shows
similar agreement with the numerically obtained results.

The rest of this paper is organized as follows. In Sec.~\ref{numeric}, we 
describe our numerical results for the spatial survival probabilities.
The analytic calculations with and without the ``zero-area'' constraint
are described in detail in Sec.~\ref{analytic}. In this section, we also
present a comparison of the analytic results for the survival probabilities 
with the numerical results presented in Sec.~\ref{numeric}.
Sec.~\ref{concl} contains a summary of the main results and a few 
concluding remarks. Some details of the steady-state properties of finite 1d EW
interfaces with periodic boundary conditions are presented in the Appendix.

\section{Numerical results}
\label{numeric}

In the numerical work, we consider a spatially discretized dimensionless form
of the 1d EW equation defined in Eq.~(\ref{eweq}). The height variable is
defined on a 1d lattice of unit spacing with periodic boundary conditions. 
Let $h_i$ be the height at lattice site $i$ with $i=1,2,\ldots,L$. The 
time-dependence of the height variables is given by
\begin{equation}
\frac{d h_i(t)}{dt} = [h_{i+1}(t)-2h_i(t)+h_{i-1}(t)] + \eta_i(t), \label{ew2}
\end{equation}
where the $\eta_i(t)$'s represent uncorrelated Gaussian noise with
$\langle \eta_i(t) \rangle = 0$ and $\langle \eta_i(t) \eta_j(t^\prime) \rangle
= 2 \delta_{ij} \delta (t-t^\prime)$. 
The Eq.~(\ref{ew2}) is thus a discretized version of the 
continuum EW equation (\ref{eweq})
with the choice $\Gamma=1$ and $D^{\prime}=1$.
These equations
are integrated forward in time  
using the simple Euler method~\cite{num_rec}. Thus, we write Eq.~(\ref{ew2}) as 
\begin{equation}
\label{discrew}
h_i(t+\delta t)-h_i(t)=\delta t [ h_{i+1}(t)
-2 h_i(t)+ h_{i-1}(t)] + \sqrt{\delta t}\, r_i(t),
\end{equation}
where each $r_i(t)$ is an independent Gaussian random number of zero mean and
variance equal to $2$. 
We use a value of $\delta t$ small enough
(i.e. $\delta t=  0.01$) to avoid any numerical instability. Steady-state
interface profiles are generated by carrying out the integration from flat
initial states ($h_i = 0$ for all $i$) to sufficiently long times (much longer 
than the time at which the width of the interface saturates). As mentioned
in Sec.\ref{intro}, we always subtract the instantaneous spatial average
of the height variables from the individual variables $\{h_i\}$, so that
the condition $\sum_{i=1}^L h_i =0$ is always satisfied. 

The steady-state spatial survival probability $Q(x)$
is measured  
as the probability
that the interface height variable does not cross zero
as one moves along the interface from an initial
point $x_0$ to the point $x_0+x$  
(since the height variables are defined on a lattice of unit spacing, both
$x_0$ and $x$ are integers between 0 and $L$.)
This probability is averaged over all initial points $x_0$ in a
steady-state configuration and also over many (typically $10^4$) 
independent realizations
of the stochastic evolution that generates the steady-state profiles.
The minimum value of the sampling interval 
$\delta x$ used in the measurement of $Q(x)$ is obviously the
lattice spacing which is equal to unity. However, it is also possible to
use a larger sampling interval, equal to a positive integer $m$, in the 
measurement of $Q(x)$ -- this is done by considering only the heights at
the lattice sites $i=km, k=1,2,\ldots$ while checking whether the height
crosses zero between the points $x_0$ and $x_0+x$. The measured survival
probability exhibits a weak dependence on the value of $\delta x$. The
FIC survival probability is measured in a similar way, except that the
initial points $x_0$ are chosen to be only those at which the height
lies between $-w$ and $+w$, with $w \ll W(L)$, the steady-state width of
the interface.   

\begin{figure}
\includegraphics[height=8.0cm,width=10.0cm,angle=0]{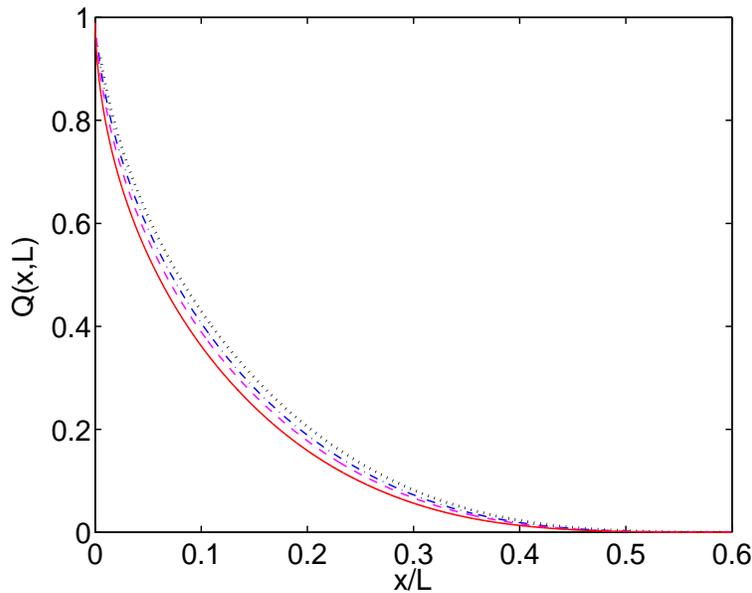}
\caption{\label{fig1} (Color online)
Plots of the spatial survival probability $Q(x,L)$ as a function of $x/L$ ($L$
is the sample size) for $L = 200$ ((black) dotted line), $L=400$ ((blue) 
dash-dotted line), $L=800$ ((magenta) dashed line), and $L=10^4$ 
((red) full line).
The same sampling interval, $\delta x=1$, is used in all cases.}
\end{figure}

\begin{figure}
\includegraphics[height=8.0cm,width=10.0cm,angle=0]{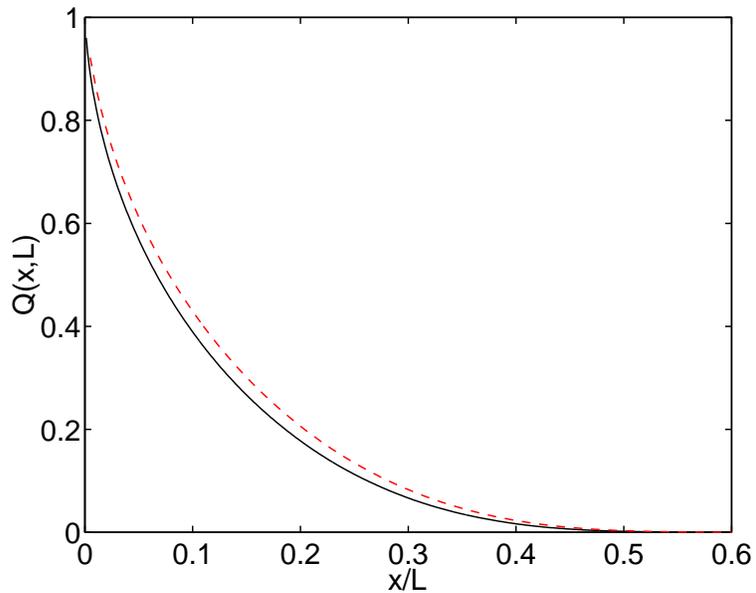} 
\caption{\label{fig2} (Color online)
Dependence of the spatial survival probability on the sampling
interval $\delta x$ used in its measurement. Plots of $Q(x,L)$ versus $x/L$ 
($L$ is the sample size) are shown for $L$ = 800, and two values of
$\delta x$: $\delta x =1$ ((black) full line) and 
$\delta x=4$ ((red) dashed line).}
\end{figure}

\begin{figure}
\includegraphics[height=8.0cm,width=10.0cm,angle=0]{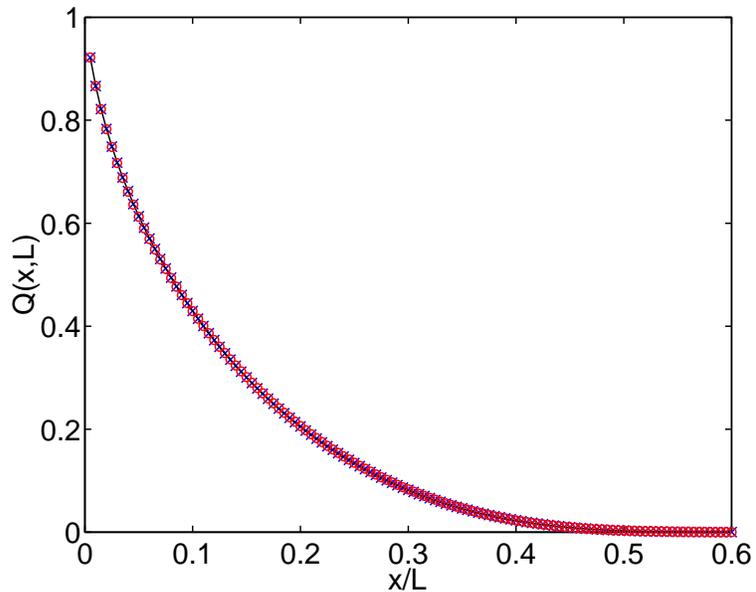} 
\caption{\label{fig3} (Color online)
The scaling behavior (Eq.~(\ref{scale1})) of the dependence of the spatial
survival probability on the sample size $L$ and the sampling interval
$\delta x$. Plots of $Q(x,L)$ versus $x/L$ are shown for three different 
sets of values of $L$ and $\delta x$ 
with $\delta x/L$ held constant: $L=200$ and $\delta x =1$ ((blue) crosses); 
$L=400$ and
$\delta x=2$ ((black) line); $L=800$ and $\delta x=4$ ((red) circles).}
\end{figure}

\begin{figure}
\includegraphics[height=8.0cm,width=10.0cm,angle=0]{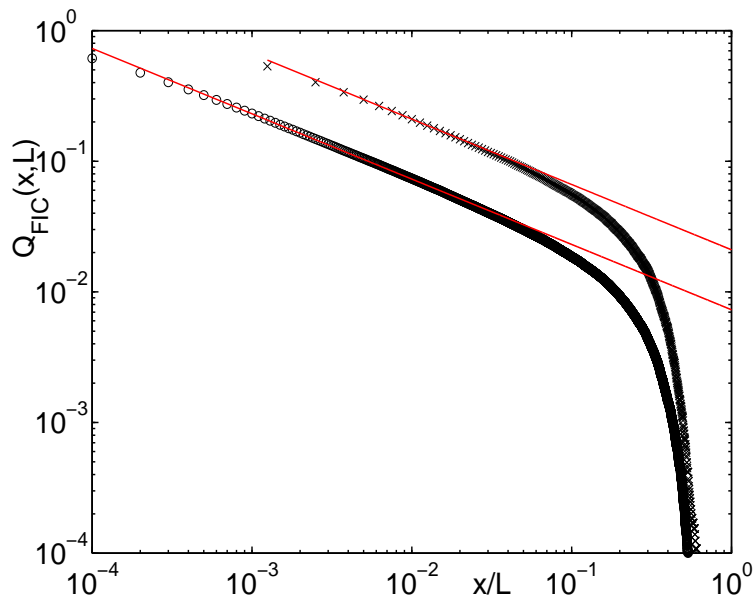} 
\caption{\label{fig4} (Color online) Double-log
plots of the FIC spatial survival probability $Q_{FIC}(x,L)$ versus $x/L$ 
($L$ is the sample size) for $L$ = 800 (crosses) 
and $L=10^4$ (circles). Fits of the initial decay of $Q_{FIC}$ to a 
power-law 
with exponent $1/2$ ($Q_{FIC}(x) \propto x^{-0.5}$) are shown by the (red)
solid lines.}
\end{figure}

Typical results for the survival probability $Q(x)$ are shown in Figs.
\ref{fig1}-\ref{fig4}. In Fig.~\ref{fig1}, we show plots of $Q(x)$ as a 
function of $x/L$ for $L=200$, 400 and 800. The values of $Q(x)$ shown in 
these plots were obtained using $\delta x=1$ (unless mentioned otherwise, all
the results shown here 
were obtained using this ``default'' value of $\delta x$).
It is clear from the plots that $Q(x)$ decreases from 1 to a value close to
0  as $x/L$ is increased 
from 0 to about 0.6. The numerical results show a weak dependence on the 
value of $L$. As we shall see shortly, this dependence arises from the use of
the same $\delta x$ in all the measurements for the different values of $L$.

The numerical calculations can not be extended to much larger values of $L$
because the time required to reach the steady state from a flat initial state
increases with $L$ as $L^z$ with $z=2$. However, we have found a different way 
of generating steady-state interface profiles for much larger values of $L$.
It is easy to show that in the steady state of the model defined in 
Eq.~(\ref{ew2}), the height difference variables $s_i \equiv h_{i+1}-h_i$,
$i=1,2,\ldots,L$ (with $h_{L+1}=h_1$ due to periodic boundary conditions)
are independent Gaussian random variables with zero mean and variance equal
to unity, apart from the obvious constraint, $\sum_{i=1}^L s_i = 0$, arising
from periodic boundary conditions. Therefore, a realization of the 
steady-state interface profile for a system of size $L$ 
may be obtained by numerically generating $L$
Gaussian random variables with the statistics mentioned above,
identifying these random variables with the $s_i$'s, and then
calculating the heights $h_i$ (with their spatial average subtracted off) in
terms of these $s_i$'s. We have calculated the spatial survival probabilities
for steady-state EW interfaces with $L=10^4$ generated this way, averaging
over 4000 independent realizations. The results for $Q(x)$ obtained from this
calculation are also shown in Fig.\ref{fig1}. It is clear that these results
for $L=10^4$ are consistent with the trend shown by the other results 
obtained from steady-state interfaces generated by numerical integration. 

In Fig.\ref{fig2}, we have shown plots of $Q(x)$ for $L=800$, obtained from
two different calculations, one with $\delta x =1$ and the other with
$\delta x=4$. The two curves are clearly different, indicating that the
measured $Q(x)$ depends weakly on the value of $\delta x$ used in the
measurement. We have found, in analogy with the results reported in
Ref.\cite{sp2} for spatial persistence probabilities, that the dependence of
$Q$ on $x$, $L$, and $\delta x$ satisfies the scaling equation
\begin{equation}
Q(x,L,\delta x) = f_d(x/L,\delta x/L), \label{scale1}
\end{equation}
where the subscript ``d'' of the scaling function is meant to indicate that here
we are considering survival probabilities measured using discrete sampling
with a finite sampling interval $\delta x$. This scaling equation implies that
plots of $Q(x)$ vs. $x/L$ for samples with different $L$ would all collapse to
the same curve if the survival probabilities for different $L$ are measured
using different values of $\delta x$, such that $\delta x/L$ is held constant.
This scaling behavior is illustrated in  Fig.\ref{fig3}.
In this Figure, we have shown plots of $Q$ vs. $x/L$ for 
$L=200$ measured with $\delta x=1$, $L=400$ measured with $\delta x=2$, and
$L=800$ measured with $\delta x=4$ (so that $\delta x/L = 1/200$ in all 3
cases). The 3 sets of data are found to collapse to the same curve, thereby
establishing the validity of the scaling form of Eq.~(\ref{scale1}). The
$L$-dependence of the results shown in Fig.\ref{fig1} may, therefore, be 
thought of as representing the dependence of the scaling function $f_d$ on
the value of its second argument, $\delta x/L$.  

Numerical results for the FIC survival probability $Q_{FIC}$ were reported
in Ref.\cite{sp2} where it was shown that it exhibits the following scaling
behavior:
\begin{equation}
Q_{FIC}(x,L,\delta x,w) = f_{FIC}(x/L, \delta x/L, w/W(L)). \label{scale2}
\end{equation}
For the sake of completeness, we have shown in Fig.\ref{fig4} our numerical
results for $Q_{FIC}$ obtained for $w/W(L)=0.02$. Two sets of data are shown,
one for $L=800$, obtained from steady-state interfaces generated by 
numerical integration, and the other 
for $L=10^4$, obtained from interfaces generated using 
Gaussian random variables,
as outlined above. In both cases, the initial decay of $Q_{FIC}$ can be
well-represented by a power law, $Q_{FIC}(x) \propto x^{-\theta_{FIC}}$ with
$\theta_{FIC}=0.5$. As in the case of the steady-state survival probability,
the dependence of $Q_{FIC}$ on the value of $L$ may be thought of as 
representing the dependence of the scaling function $f_{FIC}$ of 
Eq.~(\ref{scale2}) on the argument $\delta x/L$.

\section{Analytic calculations}
\label{analytic}

In this section, we describe in detail our analytic calculation of the spatial
survival probabilities in the stationary state of the 1d EW equation
(\ref{eweq}). We consider periodic boundary condition, $h(0,t)=h(L,t)$. The height 
can then be decomposed into a Fourier series, $h(x,t)= \sum_k {\tilde
h}(k,t)e^{i kx}$ where $k=2\pi m/L$ with $m=0,\pm 1,\pm 2 \dots$. Substituting 
this in Eq.~(\ref{eweq}) one finds that different nonzero Fourier modes decouple from each other.
This enables an exact calculation of any two-point correlation function. For example,
as shown in the Appendix, one finds for any $k\ne 0$, $\langle {\tilde h}(k,t){\tilde h}(k',t)\rangle 
=[{D^{\prime}/{\Gamma L k^2}}]\, \delta_{k+k',0}$ in the stationary limit $t\to \infty$. 
Note that the $k=0$ mode is identically zero at all times, ${\tilde h}(0,t)=0$, which
follows from the sum rule $\int_0^{L} h(x,t)dx=0$ as the height $h(x,t)$, by definition,
is always measured with respect to its spatial average. 
Since Eq.~(\ref{eweq}) is linear, the
height field $h(x,t)$ is Gaussian for all $x$ and all $t$. Using the result for the
two-point correlator mentioned above, one can then easily write down the joint probability
distribution of the Fourier modes in the stationary state,
\begin{equation}
P\left[\{{\tilde h}(k)\}\right] \propto \exp\left[- \frac{\Gamma L}{2D^{\prime}}\sum_{k} k^2 {\tilde 
h}(k){\tilde h}(-k)\right] \delta\left({\tilde h}(0)\right)
\label{jdis1}
\end{equation}
where the delta function on the right-hand side of Eq.~(\ref{jdis1}) takes care of the ``zero-area''
constraint. In terms of the actual height field $h(x,t)$, the stationary joint distribution
becomes\cite{MC1,MC2}
\begin{equation}
P_{\rm st}\left[\{ h(x)\}\right]= 2\sqrt{\pi D} L^{3/2}\, 
\exp\left[ - {1\over {4D}} \int_0^{L}dx {\left(\frac{dh}{dx}\right)}^2 \right]\,
\delta\left[h(0)-h(L)\right]\, \delta\left( \int_0^{L} h(x)dx\right),
\label{jdis2}
\end{equation}
where $D=D^{\prime}/{2\Gamma}$ and the normalization constant $2\sqrt{\pi D} L^{3/2}$,
ensuring that the joint distribution is normalized, can be
calculated explicitly\cite{MC2}.
Two delta functions in Eq.~(\ref{jdis2}) take care respectively of the periodic boundary condition
$h(0)=h(L)$ and the zero-area constraint. The stationary height distribution at any fixed point $x$
in space is, by translational invariance, independent of $x$ and is also a Gaussian
\begin{equation}    
P{\rm st}(h) = \frac{1}{\sqrt{2\pi \langle h^2\rangle}}\, e^{-h^2/{2 \langle h^2\rangle}}
\label{sshd}
\end{equation}
where the variance $\langle h^2\rangle = D^{\prime} L/{12 \Gamma}= DL/6$ can be easily
computed (see the Appendix). 

Note that in the standard literature on interfaces, one often ignores the ``zero area" 
constraint in the stationary measure\cite{Review}. This is justified if one is
interested in calculating physical properties in an infinite ($L\to \infty$) system 
where the zero mode $\tilde h(k=0)$ does not play any important role. Besides, in
the calculation of certain observables even in a finite system, such as the average width
in the stationary state or the distribution of the square of the spatially averaged width\cite{Racz},
the $k=0$ mode drops out of the calculation. However, as pointed out in Refs. \cite{MC1,MC2}, the
``zero area" constraint certainly plays a very crucial role in the calculation
of, for example, the distribution of the maximal height of the interfaces in 
a finite system. We will see below that the ``zero area" constraint 
does indeed play an important role also in the calculation of spatial
survival probabilities in a finite system. 

From the expression of the stationary measure in Eq.~(\ref{jdis2}) it is evident that
stationary paths are locally Brownian, i.e., evolve in space as, $dh(x)/dx= \xi(x)$, where 
$\xi(x)$ is a
Gaussian white noise with zero mean and a correlator,
$\langle \xi(x)\xi(x')\rangle = 2D\delta(x-x')$. For the periodic boundary condition, the stationary
path in space is, in fact, a Brownian bridge over $x\in [0,L]$ that starts at $h_0=h(0)$ at $x=0$
and ends up at the same point $h(L)=h_0$ at $x=L$. In addition, this Brownian bridge has
a total `zero area' under it. It turns out to be convenient to perform the calculations 
using the standard notations of a `temporal' Brownian motion, $dx/dt= \xi(t)$ with
$\langle \xi(t)\rangle=0$ and $\langle \xi(t)\xi(t')\rangle = 2D\delta(t-t')$. 
At the end of the calculations,
one can translate back the results to the interface problem upon identifying
(i) the height of the interface with the position of the `temporal' Brownian motion, i.e.,
$h\equiv x$ (ii) the space in the interface problem with the time in the `temporal' Brownian 
motion, i.e., $x\equiv t$. Thus, in this notation, the `temporal' Brownian bridge
starts at the initial position $x_0$ ($\equiv h_0$) at $t=0$ and ends at the same 
position $x_0$ after a time interval $t=T$ ($T\equiv L$), and enclosing under it
a total `zero area', i.e., $\int_0^T x(t) dt = 0$.    

With these notations set up, we 
first describe a calculation of the spatial survival probability in which the ``zero-area'' constraint is 
not taken into account.
Although the survival probability can be calculated exactly in this case, the
resulting analytic expression does not show good agreement with numerical
results, implying that the ``zero-area'' constraint is crucial for a correct
description of the first-passage properties. We then show that the 
``zero-area'' constraint can be taken into account in a formally 
exact path integral treatment. This treatment, however, does not lead to a 
simple closed-form expression for the survival probability that can be compared
with numerical results.  We then use a ``deterministic'' approximation
to obtain closed-form expressions for the survival probabilities
and show that the analytic results obtained this way provide a reasonably
good account of the numerical results. Note that since the stationary measure
of the 1d KPZ interface for the periodic boundary condition is the same
as that in the EW interface\cite{Review}, all our steady state results will be valid for
the 1d KPZ interface as well.    

\subsection{ Survival Probability without the Zero-area Constraint}

Let us first recall some basic results for the ordinary free `temporal' Brownian motion.
Consider a Brownian motion,
\begin{equation}
{{dx}\over {dt}} = \xi(t),
\label{brown1}
\end{equation}
where $\xi(t)$ is a zero mean Gaussian white noise with correlator $\langle 
\xi(t)\xi(t')\rangle=2D \delta(t-t')$. The free propagator of the Brownian motion
$G(x,t|x_0,t_0)$ defined as the probability that the particle will reach $x$
at time $t$ starting from $x_0$ at $t_0$ can be easily obtained by solving the
Fokker-Planck equation,
\begin{equation}
\partial_t G(x,t|x_0,t_0) = D \partial_x^2 G(x,t|x_0,t_0),
\label{prop1}
\end{equation}
subject to the initial condition, $G(x,t_0|x_0,t_0)= \delta(x-x_0)$ and the boundary conditions
that $G\to 0$ as $x\to \pm \infty$. The well known solution is given by,
\begin{equation}
G(x,t|x_0,t_0)= {1\over {\sqrt{ 4\pi D (t-t_0)}}} e^{-(x-x_0)^2/{4D(t-t_0)}},
\label{prop2}
\end{equation}
valid for all $t$ and $t_0$ and $x$ and $x_0$. We now ask: what is the probability
that the particle reaches $x$ at time $t$, starting at $x_0$ at $t_0$, but without having
crossed the zero in between? This probability $P(x,t|x_0,t_0)$ can be easily calculated by solving
the same Fokker-Planck equation, but now adding an absorbing boundary condition at $x=0$,
i.e., insisting that $P(0,t|x_0,t_0)=0$ for all $t$\cite{Redner}. The solution can be easily obtained
by the image method and is given by,
\begin{equation}
P(x,t|x_0,t_0)= {1\over {\sqrt{ 4\pi D (t-t_0)}}}\left[ e^{-(x-x_0)^2/{4D(t-t_0)}}- 
e^{-(x+x_0)^2/{4D(t-t_0)}}\right].
\label{prop3}
\end{equation}
Evidently, this solution satisfies the absorbing boundary condition at $x=0$.

Now, let us consider a Brownian bridge over the interval $[0,T]$. This means a Brownian 
motion that starts at $x_0$ at $t=0$ and ends up at the same point $x_0$ at time $T$.
Let us ask: what is the probability $Q(t,T|x_0)$ that this process (conditioned to be at $x_0$ at 
the two endpoints) does not cross zero in the interval $[0,t]$ where $0\le t\le T$? To calculate 
this probability, let us divide a typical path of the process over two intervals: $[0,t]$ and
$[t, T]$. Over the first interval $[0,t]$, a typical path starts at $x_0$ at the left end of the 
interval and lands up, say at $x$ (where $x$ is a variable) at $t$, without having crossed
the zero over the interval $[0,t]$. This probability for the left interval is (using Eq.~ 
(\ref{prop3})),
\begin{equation}
P_L(x,t|x_0,0)={1\over {\sqrt{ 4\pi D t}}}\left[ e^{-(x-x_0)^2/{4Dt}}-
e^{-(x+x_0)^2/{4Dt}}\right].
\label{propl}
\end{equation} 
Over the second interval $[t,T]$, the path starting
at $x$ at $t$ (left end of the interval $[t,T]$) reaches at $x_0$ at $T$, but there is no 
restriction over this second interval (the path is allowed to cross zero in this second interval).
Thus, this probability for the right interval is obtained from the free propagator in Eq.~ 
(\ref{prop2}),
\begin{equation}
P_R(x_0,T|x,t)={1\over {\sqrt{4 \pi D (T-t)}}}\, e^{-(x_0-x)^2/{4D(T-t)}}.
\label{propr}
\end{equation}
Due to the Markovian property of the walk, the left and the right intervals are independent.
Hence the net probability is just the product of the two probabilities, integrated over
the position $x$ at the intermediate point $t$ over $x\in [0,\infty]$. 
This will be the total probability that a path starting at $x_0$ at $t=0$ will end up
at $x_0$ at time $T$, without having crossed the zero in the interval $[0,t]$. To get the 
conditional probability $Q(t,T|x_0)$ (conditioned that the two ends are already given to be 
at $x_0$), we need to divide this probability by 
the factor $1/\sqrt{4\pi DT}$ which is just the probability of a free path landing up at $x_0$  
at $T$, starting at $x_0$ at time $0$. Thus, finally, we get
\begin{equation}
Q(t,T|x_0)= \sqrt{{T}\over { 4\pi D t(T-t)}}\int_0^{\infty} dx \left[ e^{-(x-x_0)^2/{4Dt}}-
e^{-(x+x_0)^2/{4Dt}}\right]\, e^{-(x_0-x)^2/{4D(T-t)}}.
\label{netp1}
\end{equation}
This integral can be done in closed form and one gets,
\begin{equation}
Q(t,T|x_0)= {1\over {2}}\left[1+ {\rm erf}\left(x_0 \sqrt{{T}\over {4Dt(T-t)}}\right)
-e^{-x_0^2/DT}{\rm erfc}\left( x_0{ {(T-2t)}\over { \sqrt{4DtT(T-t)} }}\right)\right],
\label{netp2}
\end{equation}
where ${\rm erf}(x)= {2\over {\sqrt \pi}}\int_0^{x} e^{-u^2}du$ is the error function
and ${\rm erfc}(x)=1-{\rm erf}(x)$.

So, now we can interpret these results in terms of the stationary state of the EW interfaces.
Identifying $x_0\equiv h_0$ and $T\equiv L$ and $t\equiv x$, $Q(x,L|h_0)$ in Eq.~(\ref{netp2})
is just the
probability that the stationary interface, given its height $h_0$ at the two ends of the sample,
does not cross zero in the spatial interval $[0,x]$ and is given by, for $h_0>0$,
\begin{equation}
Q(x,L|h_0)= {1\over {2}}\left[1+ {\rm erf}\left(h_0 \sqrt{{L}\over {4\,D\,x\,(L-x)}}\right)
-e^{-h_0^2/DL}{\rm erfc}\left( h_0{ {(L-2\,x)}\over {\sqrt{4\,D\,x\,L\,(L-x)}} }\right)\right].
\label{netp3}
\end{equation}

Now, we need to average $Q(x,L|h_0)$ over
the stationary distributions of $h_0$ given in Eq.~(\ref{sshd}), 
\begin{equation}
Q(x,L) =2\int_0^{\infty} dh_0\, Q(x,L|h_0)\, P_{\rm st}(h_0).
\label{f1}
\end{equation}
The factor $2$ comes from the fact that the stationary distribution becomes twice
its value if restricted over only the positive half-space $h_0\in [0,\infty]$.
The integral in Eq.~(\ref{f1}) can be done in closed form. We need to use the 
identity,
\begin{equation}
\int_0^{\infty} dx \,e^{-x^2} {\rm erf}(zx)= {1\over {\pi}} {\tan}^{-1}(z),
\label{iden1}
\end{equation}
which can be easily proved by differentiating both sides with respect to $z$, performing the 
resulting integral and then integrating back with respect to $z$. Our final result is:
$Q(x,L) = f(x/L)$ for all $x$ and $L$, where the scaling function $f(u)$ is given exactly by,
\begin{equation}
f(u) = {1\over {2}}\left(1- {\sqrt{3}\over {2}}\right)+ {1\over {\pi}} {\tan}^{-1}\left[
{1\over { 2\sqrt{3}}} {1\over {\sqrt{u(1-u)}}}\right] + {\sqrt{3}\over {2\pi}}{\tan}^{-1}
\left[{1\over {4}}\, {{(1-2u)}\over {\sqrt{u(1-u)}}}\right].
\label{f3}
\end{equation}
One can easily check that $f(0)=1$ as it should be. Also, note that $f(1)= 1-\sqrt{3}/2$
is nonzero. Interestingly, this scaling function $f(u)$ does not depend on the
system parameters $D^{\prime}$ or $\Gamma$.

\begin{figure}
\includegraphics[height=8.0cm,width=10.0cm,angle=0]{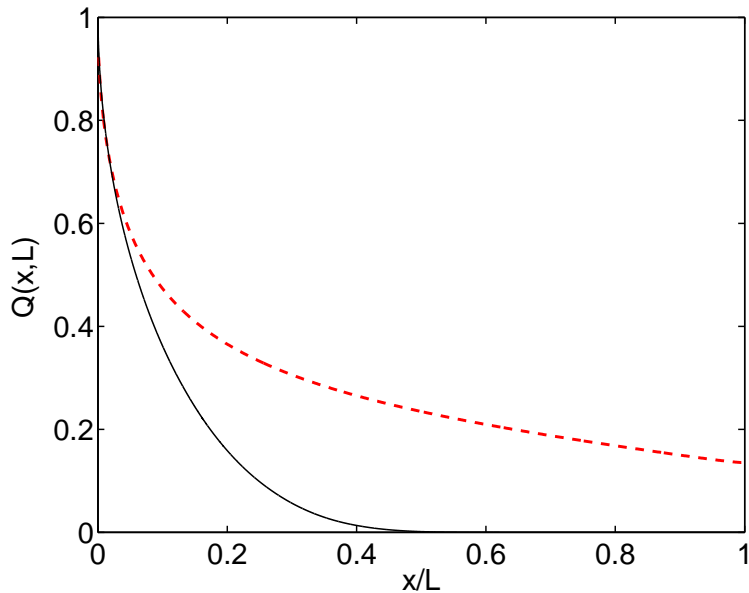}
\caption{(Color online) 
Comparison of the analytic result for the spatial survival 
probability (Eq.~(\ref{f3}) with $u=x/L$, $Q(x,L)=f(u)$), 
obtained without enforcing the zero-area constraint ((red) dashed line)
with the 
numerical result for $L=10^4$ ((black) solid line).}
\label{fig5}
\end{figure}

The function $f(u)$ in Eq.~(\ref{f3}) is plotted vs. $u$ in Fig.\ref{fig5}, and
compared with the numerical result obtained for $L=10^4$. The agreement 
between the analytic and numerical results is not satisfactory. In particular,
the numerical result for $f(u)$ goes to very small values as $u$ is increased
above 0.5, while the analytic curve shows a finite value $f(1)= 1-\sqrt{3}/2$
even at $u=1$. It is,
therefore, clear that the ``zero-area" constraint has to be included in the 
calculation for a correct description of the survival probability.

\subsection{ Survival Probability with the ``zero-area" Constraint}

In the previous section, we did not take into account the constraint that the total area under
the Brownian bridge $x(\tau)$ going from $x_0$ at time $\tau=0$ to $x_0$ at time $\tau=T$
is actually zero. In this subsection, we perform the calculation taking into
account this ``zero-area" constraint. 
\begin{figure}
\includegraphics[height=8.0cm,width=10.0cm]{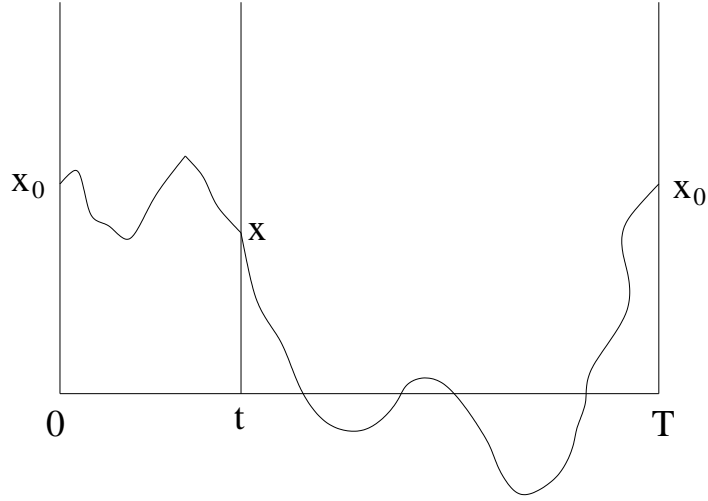}
\caption{A typical Brownian path starting at $x_0$ and reaching $x$ at time $t$ without crossing the 
origin and then freely propagating from $x$ at $t$ to $x_0$ at time $T$.}
\label{fig6}
\end{figure}

We first define $Q_0(t,T|x_0)$ to be the probability that the process $x(\tau)$ starting at $x_0$ at 
$\tau=0$ does not cross zero up to time $t$ (see Fig. \ref{fig6}) 
where $0\le t\le T$, {\em given that the
process ends up at $x_0$ at $\tau=T$ and that the total area under the process from $0$ to $T$
is $0$}. The subscript $0$ in $Q_0$ indicates that the fact that we are considering only those
paths (out of all possible paths starting at $x_0$) whose area is $0$. Note that the analogous quantity 
$Q(t,T|x_0)$
without the area constraint was computed in Eq.~(\ref{netp2}). Formally, one can
express this probability $Q_0(t,T|x_0)$ as follows,
\begin{equation}
Q_0(t,T|x_0)= { { \left\langle \left[\prod_{\tau=0}^{t}\theta\left(x(\tau)\right)\right] 
\delta\left[x(T)-x_0\right] \delta\left[\int_0^{T} x(\tau)\,d\tau\right]\right\rangle}\over
{\left\langle \delta\left[x(T)-x_0\right] \delta\left[\int_0^{T} x(\tau)\,d\tau\right]\right\rangle} 
},
\label{q01}
\end{equation}
where the angular brackets $\langle \cdots \rangle$ denote an average over all possible Brownian 
paths that start at $x_0$ at time $0$.
The numerator $\cal N$ in Eq.~(\ref{q01}) represents the joint probability of 3 events:
(i) the probability that the 
path does not cross zero upto $t$ (the first factor) (ii) the probability that the path
finally ends up at $x_0$ at time $\tau=T$ (the second factor) and (iii) the probability that 
the area under the
process upto time $T$ is zero (the third factor). The denominator $\cal D$ in Eq.~(\ref{q01}) represents 
the joint probability of events (ii) and (iii). So, the ratio ${\cal N}/{\cal D}$ represents
the conditional probability $Q_0(t,T|x_0)$, i.e., the fraction of paths that start
at $x_0$ and satisfy (i), (ii) and (iii), out of all paths that start at $x_0$ and
satisfy (ii) and (iii).

This numerator can further be split into two parts: (a) the probability that a path starting
at $x_0$ reaches $x$ at the intermediate time $t$ without crossing zero and enclosing an area,
say $A>0$ (the area is positive since the path does not cross zero in $[0,t]$) and (b) the subsequent 
probability that the path starting at $x$ at time $t$ reaches
$x_0$ at time $T$ and enclosing an area $-A$. This ensures that the total area upto $T$ is zero.
However, we then have to integrate over all possible values of $A>0$ and $x>0$. Thus, one can 
rewrite Eq.~(\ref{q01}) as,
\begin{equation}
Q_0(t,T|x_0)= { {\int_0^{\infty} dA \int_0^{\infty} dx\, G^{+} (x,A,t|x_0,0,0) G(x_0,-A, T-t|x,0,0)}
\over { G(x_0,0,T|x_0,0,0)} },
\label{q02}
\end{equation}
where we define $G(x,A,t|x_0,A_0,0)$ to be the probability that the joint i
two-variable Gaussian 
process $[x(t), A(t)=A_0 + \int_0^{t} x(\tau) d\tau)]$ (i.e., the Brownian curve itself
and the integral under the Brownian curve) reaches $[x,A]$ at time $t$, starting from
the initial value $[x_0,A_0]$ at time $0$. So, $G(x,A,t|x_0,A_0,0)$ is just the propagator
for this joint Gaussian process. Note that this process is free in the sense that it has no 
restriction for $x(t)$ to be only positive. Clearly, the denominator $\cal D$ in Eq.~(\ref{q01})
is just $G(x_0, 0, T|x_0,0,0)$ since, by definition, this quantity represents the probability 
the process $[x(t),A(t)]$ will reach its final value $[x_0,0]$ (note that the final area is zero)
at time $T$, starting from its initial value $[x_0,0]$ (since the initial area is zero). 
Similarly, the part (b) in the numerator is just $G(x_0,-A, T-t|x,0,0)$ since that
represents the joint probability that a path starting at $x$ at time $t$ and initial area $0$
will end at $x_0$ at time $T$ with area $-A$ (note that we have made a shift $[t,T]\to [0, T-t]$
which is allowed due to time translational invariance of Brownian motion). 
Finally we define $G^{+}(x,A,t|x_0,0,0)$ to be the joint probability that the
process $[x(t),A(t)]$ will reach $[x,A]$ at time $t$ starting from $[x_0,0]$ and
{\em without} $x$ crossing the origin in the interval $[0,t]$.

Now, the free propagator $G(x,A,t|x_0,A_0,0)$ is easy to compute. This is just the
joint bivariate Gaussian distribution of the random variable $x(t)$ and $A(t)=A_0+\int_0^{t} 
x(t')dt'$. Their correlation matrix can be easily computed and the propagator is just the exponential
of the inverse correlator. Indeed, this result is already well known, since this is just a
random acceleration problem: $dA/dt= x(t)$ and $dx/dt=\eta(t)$ which implies $d^2A/dt^2=\eta(t)$.
The result for the propagator can be found, for example, in Ref.\cite{ranacc,MC2}.
We just quote this result here,
\begin{equation}
G(x,A, t|x_0, A_0, 0)= { {\sqrt{3}}\over {2\pi D t^2}}\exp\left[-{1\over {D}}\left\{
{3\over {t^3}}(A-A_0-x t)(A-A_0-x_0 t)+ {1\over {t}}(x-x_0)^2\right\}\right].
\label{fpg}
\end{equation} 
Substituting this exact propagator in Eq.~(\ref{q02}), we get
\begin{eqnarray}
Q_0(t,T|x_0)&=& {\left({T\over {T-t}}\right)}^2 e^{3x_0^2/{DT}}\int_0^{\infty} dA\int_0^{\infty} dx\,
G^{+}(x,A,t|x_0,0,0)\times \nonumber \\
&\times & \exp\left[- {3\over {D(T-t)^3}}\left( A+ x(T-t)\right)\left( 
A+x_0(T-t)\right) 
-{1\over {D(T-t)}}(x-x_0)^2\right].
\label{q03}
\end{eqnarray}   
Note that we still have to determine the restricted propagator $G^{+}(x,A,t|x_0,0,0)$ and
then we have to average $Q(t,T|x_0)$ over the stationary distribution of $x_0$
to calculate $Q_0(t,T)= 2\int_0^{\infty}dx_0 Q_0(t,T,x_0) P_{\rm st}(x_0)$, where 
the distribution $P_{\rm st}(x_0)$ is given in Eq.~(\ref{sshd}).
With the identification
$h_0\equiv x_0$ and $L\equiv T$, one gets
\begin{equation}
P_{\rm st}(x_0)= \sqrt{3\over {\pi DT}}\, e^{-3x_0^2/{DT}},
\label{stat21}
\end{equation}
which gives
\begin{equation}
Q_0(t,T) = 2\sqrt{3\over {\pi DT}}\int_0^{\infty} dx_0\,  e^{-3x_0^2/{DT}} Q_0(t,T,x_0).
\label{q04}
\end{equation} 
Substituting the expression of $Q_0(t,T|x_0)$ from Eq.~(\ref{q03}) in Eq.~(\ref{q04}),
we find that the factor $e^{-3x_0^2/{DT}}$ cancels out and we get
a formally exact result,
\begin{eqnarray}
Q_0(t,T)&=& {4\sqrt{3}}\sqrt{t\over {T}} {\left({T\over {T-t}}\right)}^2 
\int_0^{\infty} {{dx_0}\over {\sqrt{4\pi Dt}}}\int_0^{\infty} dx \int_0^{\infty} dA\, 
G^{+}(x,A,t|x_0,0,0) \times \nonumber \\
&\times & \exp\left[- {3\over {D(T-t)^3}}\left(A+ x(T-t)\right)\left(A+x_0(T-t)\right)
-{1\over {D(T-t)}}(x-x_0)^2\right].
\label{q05}
\end{eqnarray}
So the remaining task is to evaluate the restricted propagator $G^{+}(x,A,t|x_0,0,0)$
which we do in the next two subsections.

\subsubsection{Exact Calculation of the Restricted Propagator $G^{+}$}

We note that the restricted propagator $G^{+}(x,A,t|x_0,0,0)$ is just the joint
probability that the process $x(t)$ starting at $x_0$ at time $0$ reaches $x$
at time $t$ without crossing the origin and that the area under the curve is $A$. 
Thus it is given by,
\begin{equation}
G^{+}(x,A,t|x_0,0,0)= 
\left\langle\left[\prod_{\tau=0}^{t}\theta\left(x(\tau)\right)\right]\delta\left[x(t)-x\right] 
\delta\left[\int_0^{t} x(\tau)\,d\tau -A \right]\right\rangle,
\label{egp1}
\end{equation} 
where the angular brackets $\langle \cdots \rangle$ denote an average over all possible
paths starting at $x_0$ at time $0$. Note that if $x_0=x=0$, this restricted
Brownian process is just a Brownian excursion over the interval $[0,t]$ 
and the propagator $G^{+}(0,A,t|0,0,0)$ is just the (unnormalized) probability density
of the area under a Brownian excursion. A Brownian excursion is simply a
Brownian path that propagates from $x(0)=0$ to $x(t)=0$ over $[0,t]$ but is
conditioned to stay positive in between. The probability distribution of 
the area under
a Brownian excursion was calculated exactly and is known
as the Airy distribution function, a complicated function that involves the
zeros of the Airy function but is not the Airy function itself\cite{Darling,Louchard,Takacs,FPV,PW}.
This Airy distribution function has been a subject of intense study for the past several years
as it has resurfaced in many problems in computer science\cite{FPV,Takacs}, graph theory\cite{MR}, and
two dimensional polygon problems\cite{Richard}. Recently it was shown that the same Airy 
distribution function
also describes the distribution of maximal height in the stationary state of fluctuating 
interfaces\cite{MC1,MC2}. In Ref. \cite{MC2}, a path integral derivation of the
area distribution under a Brownian excursion was provided. In our present problem,
we have a generalization of this problem where a Brownian path propagates from $x_0$ at $\tau=0$ to 
$x$ at $\tau=t$,
staying positive in between. The nonzero values of the initial and the final positions,
$x_0$ and $x$, make explicit calculation of the distribution of the area under
such a path difficult, as demonstrated below. Here we follow the path integral method
used in Ref. \cite{MC2} generalized to nonzero $x_0$ and $x$.  

Let us define the Laplace transform of $G^{+}$ 
with respect to the area $A$, i.e.
\begin{equation}
{\tilde G}(x,x_0,t,\lambda)= \int_0^{\infty}G^{+}(x,A,t|x_0,0,0) e^{-\lambda A}\, dA.
\label{lt1}
\end{equation}
Note that since the path is restricted to be on the positive side, the area is always positive
and hence a Laplace transform (rather than a Fourier transform) is more suitable.
Taking Laplace transform of Eq.~(\ref{egp1}) we get
\begin{equation}
{\tilde G}(x,x_0,t,\lambda)=\left\langle\left[\prod_{\tau=0}^{t}\theta\left(x(\tau)\right)\right]
\delta\left[x(t)-x\right] e^{-\lambda \int_0^t x(\tau) d\tau}\right \rangle.
\label{lt2}
\end{equation}
Using the Brownian measure of the paths in Eq.~(\ref{jdis2}), ${\tilde G}(x,x_0,t,\lambda)$ can 
be expressed
as a path integral,
\begin{equation}
{\tilde G}(x,x_0,t,\lambda)\propto \int_{x(0)=x_0}^{x(t)=x}{\cal D} x(\tau)\, \theta[x(\tau)]
\exp\left[-\int_0^{t} d\tau\, \left\{
{1\over {4D}}{\left( {{dx(\tau)}\over
{d\tau}}\right)}^2 + \lambda x(\tau)\right\}\right].
\label{pi1}
\end{equation}
Using the bra-ket notation, one can re-express the right hand side of 
Eq.~(\ref{pi1}) as a quantum mechanical propagator,
\begin{equation}
{\tilde G}(x,x_0,t,\lambda)= \langle x_0|e^{-{\hat H} t}|x\rangle,
\label{prop21}
\end{equation}
where the Hamiltonian ${\hat H}$ can be written (in the position basis),
\begin{equation}
{\hat H} = - D { {d^2}\over {dx^2}} + V(x),
\label{hamil1}
\end{equation}
where the quantum potential $V(x)=\lambda x$ for $x>0$ and in addition $V(0)=\infty$
(i.e., there is a hard wall at the origin) which takes into account the fact that all paths
in Eq.~(\ref{pi1}) are restricted to be on the positive side and can not enter the region $x<0$.
Expanding the right hand side of Eq.~(\ref{prop21}) in the eigenbasis of the Hamiltonian ${\hat H}$
we get,
\begin{eqnarray}
{\tilde G}(x,x_0,t,\lambda) &= &\sum_{E} \langle x_0|E\rangle \langle E|x\rangle e^{-E t} 
\nonumber \\
&=& \sum_{E} \psi_E(x) \psi_E^*(x_0) e^{-Et},
\label{prop22}
\end{eqnarray}
where $E$ denotes the eigenvalues of $\hat H$ and the eigenfunction $\psi_E(x)$ satisfies
the Schrodinger equation in the region $x\in [0,\infty]$,
\begin{equation}
-D {{d^2}\over {dx^2}}\psi(x) + \lambda x \psi(x) = E \psi(x),
\label{se1}
\end{equation}
with the boundary conditions, $\psi(\infty)=0$ and $\psi(0)=0$, the latter reflecting the 
hard wall at the origin. Making the change of variable, $z= (\lambda/D)^{1/3}(x-E/\lambda)$, one can 
recast the Schrodinger equation as the following Airy differential equation,
\begin{equation}
{{d^2\psi}\over {dz^2}}-z \psi =0,
\label{ai1}
\end{equation}
whose general solution is given in terms of two Airy functions,
\begin{equation}
\psi(z)= B\, {\rm Ai}(z) + C\, {\rm Bi}(z),
\label{ai2}
\end{equation}
where $B$ and $C$ are arbitrary constants. The function ${\rm Bi}(z)$ diverges as $z\to \infty$
(see Ref.\cite{abst}) indicating $C=0$. Going back to the original $x$ variable, we 
have
\begin{equation}
\psi(x) = B\ {\rm Ai}\left[ \left({\lambda\over {D}}\right)^{1/3} \left(x- {E\over 
{\lambda}}\right)\right].
\label{ai3}
\end{equation}
The other boundary condition $\psi(x=0)=0$ determines the eigenvalues,
\begin{equation}
{\rm Ai}\left[- {E\over {(D\lambda^2)^{1/3}}}\right]=0.
\label{ai4}
\end{equation}
It is known (see Ref.\cite{abst}, page 446) that the Airy function ${\rm Ai}(x)$ has zeroes on the negative
$x$ axis at $x=-\alpha_i$. For example, $\alpha_1=2.33$, $\alpha_2=4.08$, $\alpha_3=5.52$,
$\alpha_4=6.78$ etc. Thus we get 
the exact eigenvalues
\begin{equation}
E_i = \alpha_i (D\lambda^2)^{1/3}.
\label{eigen1}
\end{equation}
The amplitude $B$ in Eq.~(\ref{ai3}) is determined from the normalization, $\int_0^{\infty} |\psi_E(x)|^2 
dx=1$ and we get
\begin{equation}
|B_i|^2 = {1\over { \int_0^{\infty} {\rm Ai}^2\left[(\lambda/D)^{1/3} y- \alpha_i\right]dy} }. 
\label{amp1}
\end{equation}
Finally, putting everything back in Eq.~(\ref{prop22}) we get the exact Laplace transform of the 
restricted propagator,
\begin{equation}
{\tilde G}(x,x_0,t,\lambda)= \sum_{\alpha_i} { { {\rm Ai}[(\lambda/D)^{1/3} 
x_0-\alpha_i]\,{\rm Ai}[(\lambda/D)^{1/3} x-\alpha_i]}\over { \int_0^{\infty} 
{\rm Ai}^2\left[(\lambda/D)^{1/3} 
y- \alpha_i\right]dy} } e^{-\alpha_i D^{1/3}\lambda^{2/3} t},
\label{prop23}
\end{equation}
where $-\alpha_i$'s are the zeros of the Airy function ${\rm Ai}(x)$. Formally inverting the Laplace 
transform using the Bromwitch formula, we get the exact expression for the restricted propagator,
\begin{equation}
G^{+}(x,A,t|x_0,0,0)=\int_{\lambda_0-i\infty}^{\lambda_0+i\infty} {{d\lambda}\over {2\pi i}}\,e^{\lambda 
A}\, {\tilde 
G}(x,x_0,t,\lambda),
\label{prop24}
\end{equation}
where ${\tilde G}(x,x_0,t,\lambda)$ is given by Eq.~(\ref{prop23}) and the integration is along any 
imaginary axis whose real part $\lambda_0$ must be to the
right of all singularities of the integrand. 

Substituting this restricted propagator from Eqs. (\ref{prop24}) and (\ref{prop23}) in
Eq.~(\ref{q05}) gives us the formal exact answer of the no zero crossing probability
$Q_0(t,T)$ which, one can easily check, has the scaling form $Q_0(t,T)= f(t/T)$. However
this formal expression of the scaling function $f(u)$, though exact, is hardly useful
to make comparison with the numerical data. This is because, we can not invert the Laplace transform
explicitly in Eq.~(\ref{prop24}) (even after that one needs to do the $3$ integrals over $x_0$, $x$ 
and $A$ in Eq.~(\ref{q05}) which looks hopeless!).

So, the question is: Can one find an approximate way to estimate the restricted propagator
$G^{+}(x,A,t|x_0,0,0)$ and then use this approximate result in Eq.~(\ref{q05}), do the
$3$ integrals and derive an explicit expression for the scaling function $f(u)$? In the next 
subsection, we indeed perform these steps and derive an explicit expression of $f(u)$ which,
though not exact, is expected to be quite good.

\subsubsection{ A Deterministic Approximation for the Restricted Propagator $G^{+}$}

As a first step, let us rewrite the restricted propagator $G^{+}(x,A,t|x_0,0,0)$ in Eq.~(\ref{egp1})
as follows,
\begin{eqnarray}
G^{+}(x,A,t|x_0,0,0)&=&\left[
{{\left\langle\left[\prod_{\tau=0}^{t}\theta\left(x(\tau)\right)\right]\delta\left[x(t)-x\right]
\delta\left[\int_0^{t} x(\tau)\,d\tau -A \right]\right\rangle}\over
{\left\langle\left[\prod_{\tau=0}^{t}\theta\left(x(\tau)\right)\right]\delta\left[x(t)-x\right]
\right\rangle} }
\right]\, 
{\left\langle\left[\prod_{\tau=0}^{t}\theta\left(x(\tau)\right)\right]\delta\left[x(t)-x\right]
\right\rangle} \nonumber \\
&=& W[A,x,x_0,t]\, P(x,t|x_0,0),
\label{agp1}
\end{eqnarray} 
where we just divided and multiplied the right hand side of Eq.~(\ref{egp1}) by the same
factor $P(x,t|x_0,0)= 
{\left\langle\left[\prod_{\tau=0}^{t}\theta\left(x(\tau)\right)\right]\delta\left[x(t)-x\right]
\right\rangle}$. This factor is simply the probability that the path starting at $x_0$ at time $0$
reaches $x$ at time $t$ without crossing the origin in the interval $[0,t]$ and has already been 
calculated by the image method in Eq.~(\ref{prop3}). Thus, we have,
\begin{eqnarray}
P(x,t|x_0,0)&=& 
{\left\langle\left[\prod_{\tau=0}^{t}\theta\left(x(\tau)\right)\right]\delta\left[x(t)-x\right]
\right\rangle} \nonumber \\
&=& {1\over {\sqrt{ 4\pi D t}}}\left[ e^{-(x-x_0)^2/{4Dt}}-
e^{-(x+x_0)^2/{4Dt}}\right].
\label{agp2}
\end{eqnarray}
The quantity $W(A,x, x_0,t)$ represents the expression inside the first parenthesis on the right hand
side of Eq.~(\ref{agp1}) which we can write as
\begin{equation}
W(A,x,x_0,t)={{\left\langle
\delta\left[\int_0^{t} x(\tau)\,d\tau -A \right]
\left[\prod_{\tau=0}^{t}\theta\left(x(\tau)\right)\right]\delta\left[x(t)-x\right]\right\rangle}
\over
{\left\langle\left[\prod_{\tau=0}^{t}\theta\left(x(\tau)\right)\right]\delta\left[x(t)-x\right]
\right\rangle} },
\label{agp3}
\end{equation}
which is simply the probability distribution of the the area under the process in $[0,t]$,
{\em given that the process reaches from $x_0$ to $x$ in time $t$ without crossing the origin
in between}.

So, if we can estimate this conditional area distribution $W(A,x,x_0,t)$, then knowing the exact
$P(x,t|x_0,0)$ from Eq.~(\ref{agp2}), we will be able to estimate the restricted propagator
$G^{+}(x,A,t|x_0,0,0)$ from Eq.~(\ref{agp1}). 

Note that so far we have not made any approximation. 
To estimate the conditional area distribution $W(A,x,x_0,t)$ defined in
Eq.~(\ref{agp3}), we now make a 
``deterministic'' approximation as follows. 
Note that $W(A,x,x_0,t)$ is just the fraction of 
paths that have an area $A$, amongst 
all possible paths
that go from $x_0$ to $x$ without crossing the origin in between. 
Now, the paths that go from $x_0$ at $\tau=0$ to $x$ at $\tau=t$ without crossing
the origin in between are likely to cluster around an {\em optimal} path, i.e. most of these paths
lie ``close'' to an optimal path. This optimal path is the one which has the highest
probability amongst all possible paths going from $x_0$ at $\tau=0$ to $x$ at $\tau=t$
without crossing the origin in between.  
Assuming the existence of such an optimal path $x_{\rm opt}(\tau)$, the ``deterministic''
approximation consists in writing
\begin{equation}
W(A,x_0,x,t)\approx \delta\left(A-\int_0^t x_{\rm opt}(\tau) d\tau\right).
\label{deta1}
\end{equation}
Thus, within this aproximation we ignore the fluctuations that arise from non-optimal
paths. 

The next step is to actually find the optimal path $x_{\rm opt}(\tau)$, i.e. the path 
with the highest probability, amongst all
possible paths that  
satisfy the following constraints: (i) $x(0)=x_0$; (ii) $x(t)=x$ and (iii) $x(\tau)>0$ for
all $0\le \tau\le t$, i.e. the path stays positive
in the interval $\tau \in [0,t]$.  
One knows from the principle of least actions that the 
optimal path is
the so called ``classical'' path that satisfies Newton's equation of motion.
In our problem, the action in the path integral, $S=\frac{1}{4D}\,\int_0^t
\left(\frac{dx(\tau)}{d\tau}\right)^2 d\tau$, corresponds to that of a
free particle. So, the optimal path satisfies the Newton's law for a
free partricle, namely $d^2x/d\tau^2=0$, starting from $x(0)=x_0$ and
ending at $x(t)=x$ for all $0\le \tau\le t$. The solution is trivially,
\begin{equation}
x_{\rm opt}(\tau)= = x_0 + (x-x_0)\frac{\tau}{t}.
\label{deta5}
\end{equation}
Note that this solution automatically satisfies the condition (iii) mentioned above, i.e.
it stays positive in the interval $\tau\in [0,t]$.
Now, the area under the optimal path is simply
\begin{equation}
A_{\rm opt} = \int_0^{t} x_{\rm opt}(\tau) d\tau= \frac{1}{2}[x_0+x]t .
\label{aopt1}
\end{equation}

Before proceeding to the calculation of the survival probability with this
optimal choice, it is instructive to ask how the calculated survival probability
depends on the choice of the deterministic path. In other words, within
the ``deterministic'' approximation in Eq.~(\ref{deta1}), how does the
final result vary if instead of the optimal path in Eq.~(\ref{deta5}), we use
some other paths? To test this, we actually consider a one parameter family
of paths that satisfy the constraints (i), (ii) and (iii) above. This family
of paths is characterized 
by a single parameter $\beta>0$,
\begin{equation}
x_\beta(\tau)= x_0 + (x-x_0) \left(\frac{\tau}{t}\right)^{\beta}. 
\label{deta2}
\end{equation} 
Clearly, as shown above, $\beta=1$ corresponds to the optimal path.
In the following, we will however calculate the survival probability 
for all $\beta>0$ to test how sensitive the final answer is to
the optiaml choice $\beta=1$. We will see, somewhat surprisingly, that
the final scaling function $f(u)$ characterising the survival probability
depends only very weakly on $\beta$.

The area under the deterministic path in Eq.~(\ref{deta2}) is simply
\begin{equation}
A (\beta,x_0,x,t)= \int_0^{t} x_\beta(\tau) d\tau= 
[a\, x_0 + (1-a)\, x]t\,; \quad\quad {\rm where}\quad\, 
a=\frac{\beta}{1+\beta}.  
\label{deta6}
\end{equation}
The optimal path corresponds to the choice $\beta=1$, i.e. $a=1/2$. Our approximation 
in Eq.~(\ref{deta1}) then reads
\begin{equation}
W(A,x,x_0,t)\approx \delta\left[A- (a\, x_0+ (1-a)\, x)t \right],
\label{dt4}
\end{equation}
the optimal choice being $a=1/2$.

Within this deterministic approximation, we then have an explicit expression for the restricted 
propagator [on substituting the results in Eqs. (\ref{agp2}) and (\ref{dt4}) in Eq.~(\ref{agp1})],
\begin{equation}
G^{+}(x,A,t|x_0,0,0)\approx \frac{\delta\left[A- (a\, x_0+ (1-a)\, x)t \right]}{\sqrt{ 4\pi\, Dt}} 
\,\left[ e^{-(x-x_0)^2/{4Dt}}-
e^{-(x+x_0)^2/{4Dt}}\right].
\label{arp1}
\end{equation}

\subsubsection{Explicit Expression for the Scaling Function f(u) using the Deterministic Approximation}

On substituting the expression of $G^{+}(x,A,t|x_0,0,0)$ from Eq.~(\ref{arp1}) into Eq.~(\ref{q05}),
we can do the integral over $A$ trivially, since it involves a delta function. Inside the exponential
on the right hand side of Eq.~(\ref{q05}), we have to replace $A$ by 
$(a x_0+(1-a) x)t$. This gives,
\begin{eqnarray}
Q_0(t,T)&=&{4\sqrt{3}}\sqrt{t\over {T}} {\left({T\over {T-t}}\right)}^2
\int_0^{\infty} {{dx_0}\over {\sqrt{4\pi Dt}}}\int_0^{\infty} 
{{dx}\over {\sqrt{4\pi Dt}}}\left[ e^{-(x-x_0)^2/{4Dt}}-
e^{-(x+x_0)^2/{4Dt}}\right] \times \nonumber \\
&\times & \exp\left[ -\frac{3}{D(T-t)^3} \left(a(x_0-x)\,t+x\,T\right)
\left((1-a)(x-x_0)\,t+x_0\,T\right)-\frac{1}{D(T-t)} (x-x_0)^2 \right].
\label{ap1}
\end{eqnarray}
We next define the scaling variables: $y=x_0/\sqrt{4Dt}$, $z=x/\sqrt{4Dt}$ and $u=t/T$.
Then $Q_0(t,T)$ in Eq.~(\ref{ap1}) becomes only a function of $u=t/T$, i.e.,
$Q_0(t,T)=f(t/T)$ where the scaling function $f(u)$ is given from Eq.~(\ref{ap1}),
\begin{eqnarray}
f(u)&=& {{4\sqrt{3u}}\over {\pi (1-u)^2}}\int_0^{\infty} dy\int_0^{\infty} dz 
\left[e^{-(y-z)^2}-e^{-(y+z)^2}\right] \times \nonumber \\
&\times & \exp\left[- \frac{4u}{(1-u)^3}\left\{ \left( \gamma u^2+(3a-2)u+1 \right)y^2
+ \left( \gamma u^2 +(1-3a) u +1\right)z^2
+ \left(1+u-2\gamma\, u^2\right)yz \right\}\right],
\label{ap2}
\end{eqnarray} 
where $\gamma= 1-3a + 3a^2$.
The right hand side of Eq.~(\ref{ap2}) can be reorganized as,
\begin{equation}
f(u)={{4\sqrt{3u}}\over {\pi (1-u)^2}}\left[I_1(u)-I_2(u)\right],
\label{ap3}
\end{equation}
where
\begin{eqnarray}
I_1(u)&=& \int_0^{\infty}\int_0^{\infty}dy\, dz\, \exp\left[-r(u) y^2 -s(u) z^2 +p(u) 
yz\right],\nonumber \\
I_2(u)&=& \int_0^{\infty}\int_0^{\infty}dy\, dz\, \exp\left[-r(u) y^2 -s(u) z^2 +q(u) yz\right],
\label{ap4}
\end{eqnarray}
where $r(u)= (3(1-4a+4a^2)u^3+(12a-5)u^2+u+1)/{(1-u)^3}$, 
$s(u)=(3(1-4a+4a^2)u^3 +(7-12a)u^2 +u+1)/{(1-u)^3}$,
$p(u)=2(3(1-4a+4a^2)u^3+u^2-5u+1)/{(1-u)^3}$ and 
$q(u)=-2((-5+12a-12a^2)u^3 + 5u^2-u+1)/{(1-u)^3}$. 
To do the 
double 
integrals in Eq.~(\ref{ap4}), it is convenient to scale $y= Y/\sqrt{r(u)}$ and $z=Z/\sqrt{s(u)}$.
This gives,
\begin{eqnarray}
I_1(u)&=& {{(1-u)^3}\over {\sqrt{B(u)}}}\int_0^{\infty}\int_0^{\infty} dY dZ \exp\left[-(Y^2+Z^2 +2 
A_1(u) YZ\right], \nonumber \\
I_2(u)&=& {{(1-u)^3}\over {\sqrt{B(u)}}}\int_0^{\infty}\int_0^{\infty} dY dZ \exp\left[-(Y^2+Z^2 +2
A_2(u) YZ\right],
\label{ap5}
\end{eqnarray}
where
\begin{eqnarray}
B(u) &=& (3(1-4a+4a^2)u^3+(12a-5)u^2+u+1)(3(1-4a+4a^2)u^3+(7-12a)u^2+u+1)
\nonumber \\
A_1(u) &=& - {{3(1-4a+4a^2)u^3+u^2-5u+1}\over {\sqrt{B(u)}} } \nonumber \\
A_2(u) & =& {{(-5+12a-12a^2)u^3+5u^2-u+1}\over {\sqrt{B(u)}}}.
\label{ap6}
\end{eqnarray}
We next use the identity,
\begin{equation}
\int_0^{\infty}\int_0^{\infty} dY dZ \exp\left[-(Y^2+Z^2 +2A YZ)\right]
= {1\over {4}}\int_0^{\pi}{{d\theta}\over {1+ A \sin (\theta)}},
\label{ap7}
\end{equation}
which can be easily established by going to the polar coordinates $Y=R\cos (\theta)$
and $Z= R\sin(\theta)$ and by performing the radial integration. The integral
$J(A)=\int_0^{\pi}{{d\theta}\over {1+ A \sin (\theta)}}$ can be done in closed form
and one gets,
\begin{eqnarray}
J(A) &=&  {1\over {\sqrt{1-A^2}}}[\pi-2\, {\sin}^{-1}(A)], \,\,\,\,\,
\hbox{for}\, |A| <1, \nonumber \\
&=& {1\over {\sqrt{A^2-1}}}\log\left[{{A+\sqrt{A^2-1}}\over {A-\sqrt{A^2-1}}}\right] 
\,\,\,\,\hbox{for}\, |A| >1.
\label{ia1}
\end{eqnarray}

\begin{figure}
\includegraphics[height=8.0cm,width=10.0cm,angle=0]{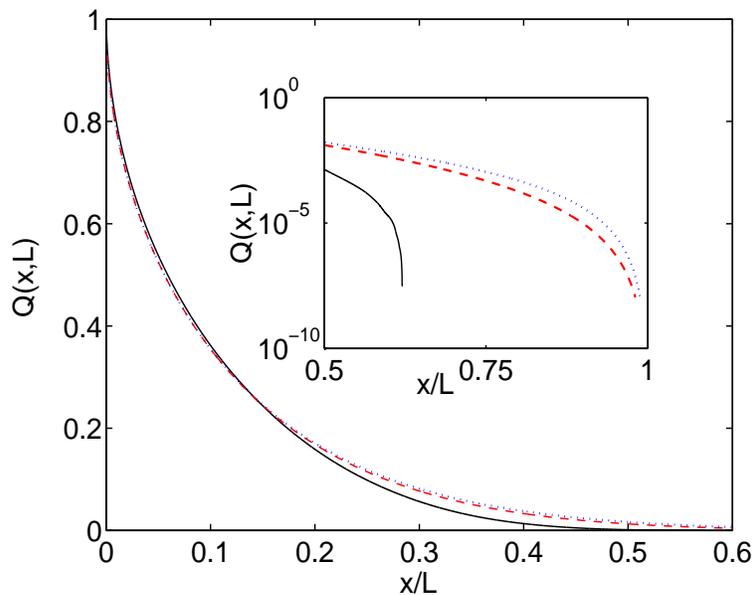}
\caption{(Color online) Comparison of the analytic result 
for the spatial survival
probability (Eq.~(\ref{fu}) with $u=x/L$, $Q(x,L)=f(u)$), obtained with the
zero-area constraint and using 
the deterministic approximation with $a=1/2$ ((red) dashed line) 
with the numerical result for $L=10^4$ ((black) solid line). The analytic 
result for $f(u)$ obtained using $a=0.1$ is also shown ((blue) dotted line)
for comparison. The inset shows the same three plots for larger values of
$u$, using a logarithmic $y$-scale to illustrate the dependence of $f(u)$ on
the parameter $a$ for relatively large values of $u$.}
\label{fig7}
\end{figure}

Putting all these results back in Eq.~(\ref{ap3}), we get an explicit result for the scaling function 
$f(u)$,
\begin{equation}
f(u)= { {\sqrt{3}} \over {\pi} } {{\sqrt{u} (1-u)}\over
{\sqrt{B(u)}}}\left[J\left(A_1(u)\right)-J\left(A_2(u)\right)\right],
\label{fu}
\end{equation}
where the functions $J$, $A_1$, $A_2$ and $B$ are given respectively in Eqs. (\ref{ia1})
and (\ref{ap6}). This is our main result, obtained using the deterministic approximation
where we keep only the contribution from the optimal path but ignore
the fluctuations around the optimal path.

The function in Eq.~(\ref{fu}) can be easily evaluated for all $0\le a\le 1$.
Note that, by definition, the $a$ dependence of $f(u)$ is symmetric about $a=1/2$, so it
suffices to use $0\le a\le 1/2$ with $a=1/2$ being the optimal choice.
In Fig. \ref{fig7} we compare the analytically obtained $f(u)$ corresponding to the
optimal choice $a=1/2$ with the actual $f(u)$ obtained via numerical
simulation on a latice of $L=10^4 $ sites.
The analytical scaling function $f(u)$ seems to compare well with
the numerical one, given especially the fact that there was no 
fitting parameter involved.

The function $f(u)$ turns out to be rather insensitive to the value of the
parameter $a$. In Fig. \ref{fig7}, we have also plotted $f(u)$ for the
choice $a=0.1$. The $f(u)$ obtained for this non-optimal choice of $a$ is
virtually indistinguishable from that obtained for the optimal choice,
$a=1/2$. This can be understood from the asymptotic behavior of $f(u)$ near
$u= 0$. 
As $u\to 0$, one can show that,
\begin{equation}
f(u) = 1- {{4\sqrt{3u}}\over {\pi}}+{8\over {\pi \sqrt{3}}} u^{3/2}+ 
{{4\sqrt{3}}\over {\pi}} u^{5/2} - \frac{32\sqrt{3}\, a(1-a)}{\pi} u^{7/2}+ (O(u^{9/2}).
\label{u0}
\end{equation}
Note that the $a$-dependence appears only in the $5$th term in the small-$u$
expansion, showing that the function $f(u)$ is highly insensitive to $a$ for
small $u$. Since $f(u)$ decreases rapidly with increasing $u$, the dependence
of $f(u)$ on $a$ for relatively large values of $u$ is not visible in the
linear plot in Fig. \ref{fig7}. This dependence is evident from the plots in
the inset of Fig. \ref{fig7} where the results for $f(u)$ for two different 
values of $a$ (the optimal value, $a=1/2$ and $a=0.1$) are shown in a
logarithmic scale and compared with the numerical data for $L=10^4$.

The dependence of $f(u)$ on the parameter $a$ for relatively large values 
of $u$ is clearly seen in the asymptotic behavior of $f(u)$ as
$u\to 1$. One finds that in powers of $\epsilon=1-u$ where $\epsilon\to 0$,
\begin{equation}
f(u=1-\epsilon)= c_4(a) \epsilon^4 
+ c_5(a)\epsilon^5 + c_6(a) \epsilon^6 + 
O(\epsilon^7),
\label{u1}
\end{equation}
where the coefficients $c_4(a)$, $c_5(a)$, $c_6(a)$ etc. are complicated functions
of the parameter $a$. In particular, for the optimal case $a=1/2$, we get
\begin{equation}
f(u=1-\epsilon)=\frac{4}{9\sqrt{3} \pi}\, \epsilon^4 + \frac{2}{3\sqrt{3} \pi}\, \epsilon^5
+ \frac{209}{270 \sqrt{3} \pi}\, \epsilon^6 + O(\epsilon^7).
\label{u1opt}
\end{equation}
 
\subsection {Survival Probability for Finite Initial Conditions}

Analytic results for the FIC spatial survival probability $Q_{FIC}$ 
discussed in 
section~\ref{intro} may be obtained from the calculations described above in
the following way. We consider the probability that, given the constraint 
that the height $h_0$ at $x = 0$
lies between $-w$ and $+w$ with $w \ll W(L)$, the steady-state width of the
interface of size $L$, the interface height does not cross zero within
distance $x$. This probability is normalized by the probability of finding
the initial height between $-w$ and $+w$ [i.e. by $2 \int_0^w P_{st}(h) dh$
where $P_{st}(h)$ is the Gaussian probability distribution for the height in
the steady state (see Eq.~(\ref{ficeq})].
In the first calculation without the zero-area constraint, we use 
Eq.~(\ref{netp3})
and integrate the quantity $Q(x,L|h_0) P_{st}(h_0)$ over $h_0$ between 0
and $w$ (with a factor 2 to take care of negative values), assuming that
$w/W(L) \ll 1$, and $u = x/L$ is of order unity. Keeping terms of the lowest (linear) order in 
$y = w/W(L)$
in the expansions of the exponentials and error functions (the latter is
justified as long as $u$ is not very close to 0 or 1), we get the following 
result
for the normalized probability that the height does not cross zero within
distance $x = u L$, when the initial height lies between $-w$ and $+w$ with $w =
y W(L)$:
\begin{equation}
Q_{FIC}(x,L,w) = \frac{y}{\sqrt{24 \pi}} \sqrt{(1-u)/u}. \label{ficeq1}
\end{equation}
This is clearly consistent with the scaling form of Eq.~(\ref{scale2}) -- the
sampling interval $\delta x$ is zero in the present continuum description. 
For small $u$, the scaling function shows a power-law decay with exponent
$1/2$, $f_{FIC}(u,0,y) \sim A(y)/\sqrt{u}$, with
$A(y) = y/\sqrt{24 \pi}$.

In the calculation with the zero-area constraint, we use the results
obtained above using the deterministic approximation. Specifically, we consider
Eq.~(\ref{ap1}) (with $t$ replaced by $x$ and $T$ replaced by $L$, so that $u
= t/T = x/L$), and do the $x_0$ integration between 0 and $w$ instead of
between 0 and $\infty$, and then divide the result by $[2 \int_0^w
P_{st}(h) dh]$ for normalization. Again, keeping terms to the lowest order
in $y = w/W(L)$, we obtain the result
\begin{equation}
Q_{FIC}(x,L,w) = f_{FIC}(u,0,y) = 
\sqrt{1/(96 \pi)}\, y \frac{(1-u)[A_2(u) - A_1(u)] r(u)} 
{\sqrt{u B(u)}}, \label{ficeq2} 
\end{equation}
where the functions $A_1(u)$, $A_2(u)$ and $B(u)$ are defined in 
Eq.~(\ref{ap6}) 
and $r(u)$ is defined in the line after Eq.~(\ref{ap4}). 
We use the optimal value, $a=1/2$, in evaluating these functions.
In the small-$u$ limit, this expression reduces to the same form,
$A(y)/\sqrt{u}$, as that found in the calculation without the zero-area 
constraint.
This is expected, since the zero-area constraint becomes important for values
of $x$ comparable to $L$.

\begin{figure}
\includegraphics[height=8.0cm,width=10.0cm,angle=0]{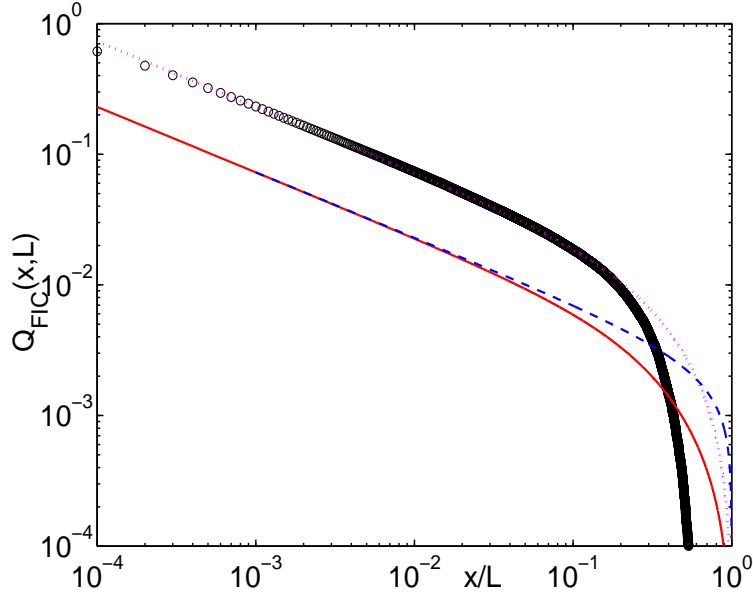}
\caption{(Color online) Comparison of analytic and numerical results for 
the FIC spatial survival probability $Q_{FIC}(x,L,w)$ with $w/W(L)=0.02$.
The (blue) dashed line and the (red) solid line show, respectively, the
analytic results obtained without the zero-area constraint
(Eq.~(\ref{ficeq1})) and with the zero-area constraint
(Eq.~(\ref{ficeq2}) with $a=1/2$). 
The numerical results for $L=10^4$ are shown by (black) circles,
and the (magenta) dotted line going through these data points represents 
the result of 
Eq.~(\ref{ficeq2}) multiplied by 3.167.}
\label{fig8}
\end{figure}

As shown in Fig.~\ref{fig8} where we have plotted the scaling functions 
according to 
Eq.~(\ref{ficeq1}) and Eq.~(\ref{ficeq2}) for $w/W(L) = 0.02$ 
versus $u=x/L$ on a log-log scale, the results obtained with and without the
zero-area constraint agree with each other for small $u$, but show
differences as $u$ increases above about 0.01. In the same Figure, we have
also shown the numerical results obtained for the same value of $y$ and
$L=10^4$. The numerical data 
show the expected $A/\sqrt{u}$ behavior for small $u$, but the
coefficient $A$ obtained from a fit to the numerical data turns out to be
substantially larger than the value $0.02/\sqrt{24 \pi} \simeq 0.0023$ 
predicted by the analytic calculation -- the value
of $A$ obtained from the fit is close to 0.0073. In Fig.\ref{fig8}, we have
also shown the result of Eq.~(\ref{ficeq2}) multiplied by 3.167, to take into
account this difference between the two values of $A$. It is clear from the
plots that the analytic result multiplied by this empirical factor provides a
good description of the numerical data -- the level of agreement is roughly 
similar to that found in Fig.\ref{fig7} for the steady-state survival 
probability.

The reason for the necessity of multiplying the analytic result by a 
numerical factor in order to obtain approximate agreement with the
numerical result lies in the use of 
discrete sampling in the numerical calculation of the FIC survival
probability. In Eqs. (\ref{ficeq1}) and (\ref{ficeq2}) above,
the survival probability goes to zero as $y \to 0$. This reflects the
fact that in the continuum limit, the probability of not crossing zero
over a finite distance $x = u L$ is zero if the initial height is zero.
This, however, is not true when the sampling of the height is done at
discrete points, $x_n = n \delta x$, where $n$ is a 
positive integer and $\delta x$ is
the sampling interval, taken to be equal to the spatial discretization
scale ($\delta x = 1$) for most of the numerical results reported here. This
is because the probability calculated in the numerical work does not take
into account the (many) zero crossings that would have taken place between
$x_1$ and $x_2$ in the continuum limit if $h(x_1)$ is very close to 0.
Even if the height at
the initial point $x_1$ is very close to zero, the probability that the
height at the next point $x_2 = x_1 + 1$ has the same sign as that
of the height at the initial point (this is the first value of the
measured ``no zero crossing probability'') is
actually close to 0.5 in all the numerical simulations -- for $h_1$ 
slightly above zero, the probability of $h_2$ remaining positive is close to
0.5 because, as discussed in section \ref{numeric}, the height difference
$s_1=h_2-h_1$ is a Gaussian random variable with zero mean. 

Our numerical results (such as those shown in Fig.\ref{fig4}) suggest that
for a fixed value of $y=w/W(L)$, the scaling functions
$f_{FIC}(u, \delta x/L,w)$ for different values of $\delta x/L$ 
differ from one another mainly by an overall numerical prefactor that
decreases as the value of $\delta x/L$ is reduced. This is why the analytic
results for the survival probability show reasonable agreement with the
numerical data (as shown in Fig.\ref{fig8}) when the former are
multiplied by a suitable numerical factor. An approximate analytic 
estimate of this
numerical factor may be obtained in the following way. Using the statistical
properties of the height difference variables $\{s_i\}$ mentioned in 
section~\ref{numeric}, it is easy to show that for $L \gg 1$, 
the probability that the 
height $h_2$ at lattice site $2$ has the same sign as that of $h_1$, the
height at site 1, is given by $0.5[1+{\rm erf}(h_1/\sqrt{2})]$. The 
``one-step'' FIC survival probability for discrete sampling with $\delta x=1$
is then given by
\begin{equation}
Q_{1d}(w,L) = Q_{FIC}(x=1,L,\delta x =1, w) =
0.5 \left[ 1+ \frac{\int_0^w dh_1 \exp(-h_1^2/2W)
{\rm erf}(h_1/\sqrt{2})}{\int_0^w dh_1 \exp(-h_1^2/2W)}\right], \label{ficeq3}
\end{equation}
where $W(L)=\sqrt{L/12}$ is the steady-state width of the interface. For
$w \ll W(L)$, this becomes a function of $w$ only:
\begin{equation}
Q_{1d}(w) = 0.5 [ 1+ {\rm erf}(w/\sqrt{2}) -\sqrt{2/\pi} (1-e^{-w^2/2})/w].
\label{ficeq4}
\end{equation}
As expected, $Q_{1d}(w)$ goes to 0.5 as $w \to 0$. Our numerical results
for $Q_{1d}(w)$ are in good agreement with this analytic prediction. 
The ``one-step'' 
FIC survival probability in the continuum limit may be obtained by setting
$x=1$ in Eq.~(\ref{netp3}), integrating $Q(x=1,L|h_0) P_{st}(h_0)$ over $h_0$ 
between 0 and $w$, and dividing by $\int_0^w P_{st}(h_0) dh_0$ for 
normalization. For $w \ll W(L)$, $L \gg 1$, this leads to the result
\begin{equation}
Q_{1c}(w) = Q_{FIC}(x=1,L,\delta x =0, w)=
{\rm erf}(w/\sqrt{2}) -\sqrt{2/\pi} (1-e^{-w^2/2})/w =
2 Q_{1d}(w) -1. \label{ficeq5}
\end{equation}
Thus, $Q_{1c}(w)$ goes to zero as $w \to 0$, as expected. However, both 
$Q_{1d}(w)$ and $Q_{1c}(w)$ approach unity for $L \gg 1$, $w \gg 1$,
$w/W(L) \to 0$, indicating that the numerical and analytic results would
agree with each other in this limit. 
For $w/W(L)=0.02$, $L=10^4$ (the values for which numerical results are 
shown in Fig.\ref{fig8}), the values of the one-step survival probabilities are
$Q_{1d}=0.612$ and $Q_{1c}=0.224$. The ratio of these two number is
2.73, which is slightly smaller than the empirical multiplicative factor used
in Fig.\ref{fig8} to bring the analytic result for the FIC survival probability
in approximate agreement with the numerical data. 
This difference reflects the fact that the empirical value
used in Fig.\ref{fig8} was obtained by considering the 
analytic and numerical results for a range of values of $x$, whereas the
analytic estimate of the multiplicative factor is obtained by considering
only one point.
\section{Summary and discussions}
\label{concl}

In this paper, we have presented analytic and numerical results for the
spatial survival probabilities for 1d EW interfaces in the steady state. 
In 1d the same steady state results also hold for the KPZ interface.
We have studied both the
steady-state and the FIC survival probabilities. The
numerical results show that these survival probabilities exhibit simple
scaling behavior as functions of the system size and the sampling interval
used in the measurement. In the analytic work, we have used a ``deterministic''
approximation to obtain closed-form expressions for these scaling functions
from an exact path integral treatment of a mapping of the problem to 1d
Brownian motion. The analytic results show fairly good agreement with the
numerical data without having to use any adjustable parameter. The
remaining differences between the analytic and numerical results may be
attributed to (a) the use of the deterministic approximation in obtaining the
analytic results, and (b) the
use of a finite sampling interval in the numerical calculations. The effect
of discrete sampling is small in the case of the steady-state survival
probability. For the FIC survival probability, the dependence of the
numerical results on the value of the sampling interval used in the 
measurement is approximately described by an overall multiplicative factor.
The value of this multiplicative factor can be estimated analytically by
considering the one-step survival probability. Further analytic work on
the dependence of the survival probabilities on the sampling interval
would be interesting and useful, especially because measurements of these 
probabilities in simulations and experiments always involve discrete
sampling. While some progress in this direction
has been made~\cite{sm2}, a complete analysis of the effects of discrete
sampling remains a challenging theoretical problem.

On the experimental side, fluctuating steps on a vicinal surface provide a
physical realization of 1d EW interfaces if the kinetics is dominated by
attachment/detachment processes~\cite{bart}. While experimental studies
of temporal persistence and survival probabilities have been carried 
out~\cite{exp1,exp2,exp3,exp4} for this system, we are not aware of any 
experimental investigation of spatial first-passage properties of 
fluctuating steps. Such investigations would be most welcome.
In other experimental systems such as combustion fronts in paper
which also involve 1d interfaces described by the EW or the KPZ equation,
the spatial persistence has been recently investigated~\cite{exp_px}.
It would be interesting to see if our theoretical predictions on the
spatial survival probability can also be verified experimentally in
such systems.



\appendix

\section{ Finite 1d EW interfaces with periodic boundary condition}

Consider the EW equation (\ref{eweq}) on a finite line of size $L$ with periodic boundary condition,
$h(x+L,t)=h(x,t))$. Since $h(x,t))$ is a periodic function, one can decompose it into a Fourier series,
$h(x,t) = \sum_k {\tilde h}(k,t) e^{ikx}$ where $k=2\pi m/L$ with $m=0,\pm 1,\pm 2,\ldots$. Thus 
\begin{equation}
h(x,t)= \sum_{m=-\infty}^{\infty} {\tilde h}(m,t)\, e^{2\pi i m x/L},
\label{a1}
\end{equation}
where the Fourier coefficients ${\tilde h}(m,t)$ are given by the inversion formula,
\begin{equation}
{\tilde h}(m,t)= {1\over {L}}\, \int_0^{L} h(x,t)\, e^{-2\pi i m x/L}\, dx.
\label{a2}
\end{equation}
Substituting Eq.~(\ref{a1}) in the EW equation (\ref{eweq}) one gets,
\begin{equation}
\partial_t {\tilde h}(m,t)= -{{4\pi^2m^2}\over {L^2}}\,\Gamma\, {\tilde h}(m,t) + {\tilde \eta}(m,t),
\label{a3}
\end{equation}
for all $m\ne 0$,
where $\langle {\tilde \eta}(m,t)\rangle =0$ and $\langle {\tilde \eta}(m,t){\tilde \eta}(m',t')=
{{2D'}\over {L}} \delta_{m+m', 0}$ with $\delta_{m,n}$ being the Kronecker delta function.
Note that for $m=0$, ${\tilde h}(0,t)=0$ at all $t$, because of the sum rule, $\int_0^{L} h(x,t)dx=0$.
Solving Eq.~(\ref{a3}) with initial condition ${\tilde h}(m,0)=0$, we get for $m\ne 0$,
\begin{equation}
{\tilde h}(m,t)= \int_0^{t} e^{-4\pi^2 m^2 \Gamma (t-t')/L^2}\, {\tilde \eta}(m,t')\,dt'.
\label{a4}
\end{equation}
Using Eq.~(\ref{a4}) and the properties of the noise, one can easily compute the two-point
equal time correlation function and we get
\begin{equation}
\langle {\tilde h}(m_1,t){\tilde h}(m_2,t)\rangle = {{2D'}\over {\Gamma L}}\, {{L^2}\over {8\pi^2 
m_1^2}}
\left[1- e^{-8\pi^2 m_1^2 \Gamma t/L^2}\right]\, \delta_{m_1+m_2, 0}.
\label{a5}
\end{equation}
This gives the two point correlation function in real space,
\begin{equation}
\langle h(x_1,t)h(x_2,t)\rangle = {{2D'}\over {\Gamma L}}\,\sum_{m_1\ne 0}{{L^2}\over {8\pi^2 
m_1^2}}\left[1- e^{-8\pi^2 m_1^2 \Gamma t/L^2}\right]\, e^{2\pi i m (x_1-x_2)/L}.
\label{a6}
\end{equation}
Note that the sum in Eq.~(\ref{a6}) runs from $m=-\infty$ to $m=\infty$ but does not include the $m=0$ 
term, since we have used 
the fact that ${\tilde h}(m=0, t)=0$.
In particular, putting $x_1=x_2=x$, we get the on-site variance, which becomes
independent of $x$ as expected, due to the translational invariance. This gives 
\begin{equation}
\langle h^2(0)\rangle =  {{2D'}\over {\Gamma L}}\,\sum_{m_1\ne 0}{{L^2}\over {8\pi^2
m_1^2}}\left[1- e^{-8\pi^2 m_1^2 \Gamma t/L^2}\right].
\label{a7}
\end{equation}
In the stationary limit $(t\to \infty)$, one gets from Eq.~(\ref{a7})
\begin{equation}
\langle h^2(0)\rangle = {{D'\, L}\over {2\pi^2 \Gamma }}\, \sum_{m=1}^{\infty} {1\over {m^2}}.
\label{a8}
\end{equation}
Using the identity, $\sum_{m=1}^{\infty} = \pi^2/6$, we get the formula for the on-site variance in the 
stationary limit
\begin{equation}
\langle h^2(0)\rangle = {D'\over {12 \Gamma}}\, L = \frac{D\, L}{6},
\label{a9}
\end{equation}
where $D=D^{\prime}/{2\Gamma}$. Since the Eq.~(\ref{eweq}) is linear, it follows then that
the single site height distribution is a pure Gaussian at all times. In particular, the
stationary distribution is given by
\begin{equation}
P_{\rm st} (h_0)= \frac{1}{\sqrt{2\pi \langle h^2(0)\rangle}}\, \exp[-h_0^2/{2\langle h^2(0)\rangle}],
\label{a10}
\end{equation}
where $\langle h^2(0)\rangle$ is given in Eq.~(\ref{a9}).


\begin{thebibliography}{99}
\bibitem{sm1} For a review on temporal persistence, see
S. N. Majumdar, Curr. Sci. {\bf 77}, 370 (1999).


\bibitem{th1} J. Krug,  H. Kallabis,  S.N.\ Majumdar, S.J.\ Cornell,
A.J.\ Bray, and C. Sire, Phys.\ Rev.\ E {\bf 56}, 2702 (1997).

\bibitem{th2} H. Kallabis and J. Krug, Europhys.\ Lett.\ {\bf 45}, 20 (1999).

\bibitem{th3} M. Constantin, C. Dasgupta, P. Punyindu Chatraphorn, Satya N.
Majumdar and S. Das Sarma, Phys. Rev. E {\bf 69}, 061608 (2004).

\bibitem{th4} C. Dasgupta, M. Constantin, S. Das Sarma and Satya N. Majumdar,
Phys. Rev. E {\bf 69}, 022101 (2004).


\bibitem{exp1} D. B. Dougherty, I. Lyubinetsky, E. D. Williams,
M. Constantin, C. Dasgupta and S. Das Sarma, Phys. Rev. Lett {\bf 89}, 
136102 (2002).

\bibitem{exp2} D. B. Dougherty, O. Bondarchuk, M. Degawa and E. D. Williams,
Surf. Sci.{\bf 527}, L213 (2003).

\bibitem{exp3} O. Bondarchuk, D. B. Dougherty, M. Degawa, E. D. Williams, M.
Constantin, C. Dasgupta and S. Das Sarma, Phys. Rev. B {\bf 71}, 045426 (2005).

\bibitem{exp4} D. B. Dougherty, C. Tao, O. Bondarchuk, W. G. Cullen,
E. D. Williams, M. Constantin, C. Dasgupta and S. Das Sarma, Phys. Rev. E
{\bf 71}, 021602 (2005).

\bibitem{sp1} S. N. Majumdar and A. J. Bray, Phys. Rev. Lett. {\bf 86},
3700 (2001).

\bibitem{sp2} M. Constantin, S. Das Sarma, and C. Dasgupta, Phys. Rev. 
E {\bf 69}, 051603 (2004).

\bibitem{kpz} M. Kardar, G. Parisi, and Y.-C. Zhang, Phys. Rev. Lett. {\bf 56}, 889 (1986).

\bibitem{exp_px} J. Merikoski, J. Maunuksela, M. Myllys, J. Timonen, and 
M. J. Alava, Phys. Rev. Lett. {\bf 90}, 024501 (2003).

\bibitem{ew} S. F. Edwards and D. R. Wilkinson, Proc. R. Socs. London,
Ser.A {\bf 381}, 17 (1982)

\bibitem{bart} N. C. Bartelt, J. L. Golding, T. L. Einstein, and E. D. 
Williams, Surf. Sci. {\bf 273}, 252 (1992).

\bibitem{sm2} S.N. Majumdar, A.J. Bray and G.C.M.A. Ehrhardt, Phys. Rev.
E {\bf 64}, 015101(R) (2001); G.C.M.A Ehrhardt, A.J. Bray and S.N. Majumdar,
Phys. Rev. E {\bf 65}, 041102 (2002).


\bibitem{num_rec} W. H. Press {\it et al.}, {\em  Numerical Recipes}
(Cambridge University Press, Cambridge, 1989).

\bibitem{MC1} S.N. Majumdar and A. Comtet, Phys. Rev. Lett. {\bf 92}, 225501 (2004).

\bibitem{MC2} S.N. Majumdar and A. Comtet, J. Stat. Phys. {\bf 119}, 777 (2005).

\bibitem{Review} A.L. Barabasi and H.E. Stanley, {\em Fractal Concepts in Surface
Growth} (Cambridge University Press, Cambridge, England, 1995); 
T. Halpin-Healy and Y.-C. Zhang, {\em  Phys. Rep.} {\bf 254} 215 (1995).
J. Krug, {\em Adv. Phys.} {\bf 46}, 139 (1997).

\bibitem{Racz} G. Foltin, K. Oerding, Z. Racz, R.L. Workman, and R.K.P. Zia,
Phys. Rev. E {\bf 50}, R639 (1994).


\bibitem{Redner} S. Redner, {\em A Guide to First-passage Processes}
(Cambridge University Press, Cambridge 2001).

\bibitem{ranacc} T.W. Burkhardt, J. Phys. A: Math. Gen.
{\bf 26}, L1157 (1993). 

\bibitem{Darling} D.A. Darling, Ann. Probab. {\bf 11}, 803 (1983).

\bibitem{Louchard} G. Louchard, J. Appl. Prob. {\bf 21}, 479 (1984).

\bibitem{Takacs} L. Takacs, Adv. Appl. Prob. {\bf 23} 557, (1991); J. Appl. Prob.
{\bf 32}, 375 (1995).

\bibitem{FPV} P. Flajolet, P. Poblete, and A. Viola, Algorithmica
{\bf 22}, 490 (1998); P. Flajolet and G. Louchard, Algorithmica
{\bf 31}, 361 (2001).

\bibitem{PW} M. Perman and J.A. Wellner, Ann. Appl. Prob. {\bf 6}, 1091 (1996).

\bibitem{MR} C.L. Mallows and J. Riordan, Bull. Am. Math. Soc.
{\bf 74}, 92 (1968); E.M. Wright, J. Graph Theory {\bf 1}, 317 (1977); 
I. Gessel, B.E. Sagan, and
Y.-N. Yeh, J. Graph Theory {\bf 19}, 435 (1995).

\bibitem{Richard} C. Richard, J. Stat. Phys. {\bf 108}, 459 (2002);
M.J. Kearney, J. Phys. A: Math. Gen., {\bf 37}, 8421 (2004). 


\bibitem{abst} M. Abramowitz and I.A. Stegun, {\em Handbook of Mathematical Functions}
(Dover, New York, 1973).



\end{thebibliography}
\end{document}